\documentclass[aps,twocolumn,superscriptaddress,pra,showpacs,nofootinbib]{revtex4-1}

\usepackage[T1]{fontenc}
\usepackage{times}
\usepackage{amsmath}
\usepackage{amssymb}
\usepackage{amsthm}
\usepackage{bbm}
\usepackage{xcolor}
\usepackage{multirow}
\usepackage{graphicx}

\usepackage{array}
\usepackage{soul}
\usepackage{tcolorbox}
\usepackage{dirtytalk}

\usepackage{tikz}

\usepackage{pgfplots}
\usetikzlibrary{graphs}
\pgfplotsset{compat=1.10}
\usepgfplotslibrary{fillbetween}
\usepackage{color}
\usepackage{setspace}
\usepackage{pifont}

\usepackage[utf8]{inputenc}

\newcommand{\bra}[1]{\langle #1 |}
\newcommand{\ket}[1]{| #1 \rangle}
\newcommand{\proj}[1]{| #1 \rangle\!\langle #1 |}
\newcommand{\GHZ}{{\rm GHZ}}

\newcommand{\B}{\mathcal{B}}

\newcommand{\I}{\mathcal{I}}
\newcommand{\M}{\mathbb{M}}
\newcommand{\Id}{\mathbb{I}}
\newcommand{\IFull}[1]{\mathcal{I}^{\rm all}_{\rm {#1}}}
\newcommand{\IPart}[1]{\mathcal{I}_{\rm {#1}}}

\newcommand{\BFull}[1]{\mathcal{B}^{\rm all}_{{\rm #1}}}
\newcommand{\BPart}[1]{\mathcal{B}_{{\rm #1}}}

\renewcommand{\S}{\mathcal{S}}
\newcommand{\Q}{\mathcal{Q}}

\newcommand{\SQmax}[1]{\S^{\Q,*}_{#1,\gamma}}

\newcommand{\ketG}{\ket{{\rm G}}}

\newcommand{\tr}{\ensuremath{\operatorname{tr}}}

\newcommand{\vecP}{\vec{P}}
\newcommand{\vecx}{\vec{x}}

\definecolor{nred}{rgb}{0.9,0.1,0.1}
\definecolor{nblack}{rgb}{0,0,0}
\definecolor{nblue}{rgb}{0.2,0.2,0.8}
\definecolor{ngreen}{rgb}{0.2,0.6,0.2}

\DeclareMathOperator*{\argmax}{argmax}
\definecolor{myurlcolor}{rgb}{0,0,0.4}
\definecolor{mycitecolor}{rgb}{0,0.5,0}
\definecolor{myrefcolor}{rgb}{0.5,0,0}
\usepackage{hyperref}

\definecolor{hotmagenta}{rgb}{1.0, 0.11, 0.81}
\usepackage[capitalize]{cleveref}

\renewcommand{\L}{\mathcal{L}}

\begin{document}
\title{Exploring Bell inequalities for the device-independent \\certification of multipartite entanglement depth}

\author{Pei-Sheng Lin}%
\affiliation{Department of Physics and Center for Quantum Frontiers of Research \& Technology (QFort), National Cheng Kung University, Tainan 701, Taiwan}
\author{Jui-Chen Hung}
\affiliation{Department of Physics, National Cheng Kung University, Tainan 701, Taiwan}
\author{Ching-Hsu Chen}
\affiliation{Department of Electrophysics, National Chiayi University, Chiayi 300, Taiwan}
\author{Yeong-Cherng Liang}
\email{ycliang@mail.ncku.edu.tw}
\affiliation{Department of Physics and Center for Quantum Frontiers of Research \& Technology (QFort), National Cheng Kung University, Tainan 701, Taiwan}

\begin{abstract}
Techniques developed for device-independent characterizations allow one to certify certain physical properties of quantum systems without assuming any knowledge of their internal workings. Such a certification, however, often relies on the employment of device-independent witnesses catered for the particular property of interest. In this work, we consider a one-parameter family of multipartite, two-setting, two-outcome Bell inequalities and demonstrate the extent to which they are suited for the device-independent certification of genuine many-body entanglement (and hence the entanglement depth) present in certain well-known multipartite quantum states, including the generalized Greenberger-Horne-Zeilinger (GHZ) states with unbalanced weights, the higher-dimensional generalizations of balanced GHZ states, and the $W$ states. As a by-product of our investigations, we have found that, in contrast with well-established results, provided trivial qubit measurements are allowed, full-correlation Bell inequalities can also be used to demonstrate the nonlocality of weakly entangled unbalanced-weight GHZ states. Besides, we also demonstrate how two-setting, two-outcome Bell inequalities can be constructed, based on the so-called GHZ paradox, to witness the entanglement depth of various graph states, including the ring graph states, the fully connected graph states, and some linear graph states, etc.
\end{abstract}
\date{\today}
\maketitle

\section{Introduction}

Entanglement~\cite{Horodecki:RMP} is a feature of quantum theory that has no analog in classical theory. Various tasks that were thought to be impossible in classical theory are now  made possible by utilizing this precious resource. For example, secure secret keys between two remote parties can be established with the help of pairs of entangled states~\cite{Ekert91}. Even an unknown quantum state can be teleported~\cite{QuantumTeleportation} from one place to another intact if one has access to maximally entangled states. These examples involve only entanglement between two parties but the concept of entanglement goes beyond this. With more complicated entanglement structures~\cite{He:PRX:2018}, such as that endowed by the two-dimensional graph states~\cite{Briegel_ClusterStates_01}, one can perform universal quantum computation~\cite{UniversalOneWayQC_2003} through a  one-way quantum computer~\cite{OneWayQC_2001}. Entanglement can thus be seen as useful resources for a variety of different tasks.

To utilize the advantages of entangled states, one must first prepare the desired entangled states relevant to the protocols. However, imperfections always arise during the preparation stage and, hence, one may end up having something very different from the desired target state. How can one be sure that the prepared state is indeed useful for the tasks in mind? Even if one only wishes to certify that the state is entangled, an analogous question remains, namely, how can one perform such a certification when the measurement devices themselves may be subjected to imperfections~\cite{Rosset:PRA:2012}? 

Tools developed for device-independent quantum information (DIQI)~\cite{Brunner14, Scarani_DIQI_12} are tailor made to address such problems. More precisely, the statistics derived from locally measuring the prepared state allow one to reach nontrivial conclusions about the system with {\em minimal} assumptions, specifically, no {\em assumption} about the internal workings of {\em any} of the devices involved is needed. For instance, whenever the measurement statistics observed in a Bell experiment show a violation of Bell inequalities~\cite{Bell64}, one can immediately conclude that the shared state is entangled~\cite{Bancal11}. Moreover, the amount of entanglement present can be lower bounded from the violation of Bell inequalities~\cite{Moroder13,SLChen16,Cavalcanti16,SLChen18}. In some cases, it is even possible to certify nontrivial properties of the measuring apparatus~\cite{Mayers04,SLChen16,Cavalcanti16,Renou:PRL:2018,Bancal:PRL:2018}.

Going beyond the bipartite scenario, the structure of entanglement can be much more complicated~\cite{He:PRX:2018}. For example,  the entanglement may not involve all subsystems but only a subset of them. In other words, an $n$-partite state that is entangled may only contain $k$-body entanglement, with $k<n$. To capture this, the notion of entanglement depth~\cite{Anders_EntanglementDepth_01}, which is closely related to that of $k$-producibility~\cite{Guhne_Produciblity_05}, has been introduced. More recently, a device-independent certification of the entanglement depth was theoretically shown to be feasible~\cite{Liang:PRL:2015} (see also Refs.~\cite{Nagata:PRL:2002,Yu:PRL:2003,Aloy:1807.06027}) for certain multipartite quantum states, such as the family of Greenberger-Horne-Zeilinger (GHZ) qubit states.

To produce a genuinely multipartite entangled (GME) quantum state (one where the entanglement indeed involves all subsystems) beyond a handful of subsystems is experimentally challenging~\cite{Monz:PRL:2011,He:PRX:2018}. Interestingly, a device-independent certification of the presence of this strongest type of entanglement remains possible~\cite{Bancal11,Moroder13}, even for an arbitrary GME pure quantum state~\cite{Zwerger_DICGME_18}. The generic construction provided in Ref.~\cite{Zwerger_DICGME_18}, however, apparently requires a Bell test that has too many measurement settings, thus making it infeasible in many practical situations.

In this work, we consider Bell inequalities that require only two binary-outcome measurements per party and characterize their ability to serve as device-independent witnesses for entanglement depth for various multipartite pure states, including unbalanced-weight qubit GHZ states, higher-dimensional generalization of the GHZ states, $W$ states~\cite{WStates_00}, one-dimensional cluster states~\cite{Briegel_ClusterStates_01} with periodic or closed boundary conditions, as well as a few two-dimensional graph states. For the first two kinds of states, we make use of the one-parameter families of witnesses proposed (but never analyzed) in Ref.~\cite{Liang:PRL:2015}. For the graph states, we construct new Bell inequalities based on  the so-called GHZ paradox~\cite{Kafatos1989-KAFBTQ, Mermin_90} and show that these inequalities can indeed be used to witness the GME nature of (many of) these states.

The rest of this paper is organized as follows. In Sec.~\ref{sec:Preliminaries} we introduce the notations used in this paper and briefly recall the definitions of producibility and entanglement depth. Then, we present our analysis of the generalized witnesses introduced in~\cite{Liang:PRL:2015}. We also show in Sec.~\ref{sec:GeneralizedGHZ} that those inequalities could be used to detect the genuine multipartite entanglement of generalized GHZ states. In Sec~\ref{sec:DIWEDforGraph} we present our inequalities constructed for graph states and show that the genuine multipartite entanglement of the states can be revealed. In Sec~\ref{sec:Conclusions} we summarize our work and discuss some possible future research directions.           

\section{Preliminaries}
\label{sec:Preliminaries}

\subsection{Notations}

The basis for employing a device-independent witness for entanglement depth (DIWED) is a multipartite Bell experiment where a shared quantum state $\rho$ is locally measured to estimate the correlation present between the observed measurement outcomes. More precisely, consider $n$ spatially separated parties and with each of them allowed to perform two dichotomic (i.e., binary-outcome) measurements. We denote the measurement choice of the $i$th party and the corresponding measurement outcome, respectively, by $x_i \in \{1, 2\}$ and $a_i \in \{+1,-1 \}$. We then use the $n$-bit vector $\vec{x} = (x_1, x_2, \cdots,x_n)$ to describe the collection of measurement settings of all parties and label their respective measurement outcomes by $\vec{a} = (a_1,a_2,\cdots,a_n)$. According to Born's rule, the probability of observing outcomes $\vec{a}$ given the measurement choices $\vec{x}$ read as:
\begin{equation}
  \label{eq:BornRule}
  P(\vec{a}|\vec{x}) = \tr\left(\rho\;M^{(1)}_{a_1|x_1}\otimes M^{(2)}_{a_2|x_2} \otimes\cdots\otimes M^{(n)}_{a_n|x_n}\right),
\end{equation}
where $\{M^{(i)}_{a_i|x_i}\}_{a_i}$ is the positive-operator-valued-measure (POVM) used to describe the $i$th party's $x_i$th measurement. Given the conditional probabilities, we can further define the $n$-partite correlators as
\begin{equation}\label{Dfn:Correlator}
  E_n(\vec{x}) = \sum_{a_1,a_2,\cdots,a_n = \pm 1} \left(\prod_{i=1}^{n}a_i\right)P(\vec{a}|\vec{x}).  
\end{equation}

In these notations, a (linear) Bell expression is defined by a certain linear combination of conditional probabilities $P(\vec{a}|\vec{x})$ with weights specified by $\beta^{\vec{x}}_{\vec{a}}$. By maximizing this linear expression over all $\vecP:=\{P(\vec{a}|\vec{x})\}_{\vec{a},\vec{x}}$ in $\L$ (the Bell-local set~\cite{Brunner14, Scarani_DIQI_12}), one obtains a Bell inequality:
\begin{equation}
  \label{eq:LocalCorrelations}
  \I_n: \sum_{\vec{a},\vec{x}}\beta_{\vec{a}}^{\vec{x}}P(\vec{a}|\vec{x}) \overset{\L}{\leq} S_n^\L=\max_{\vec{P'}\in\L} \sum_{\vec{a},\vec{x}}\beta_{\vec{a}}^{\vec{x}}P'(\vec{a}|\vec{x}).
\end{equation}
It is well-known~\cite{Werner:PRA:1989} that regardless of the measurements $\{M^{(i)}_{a_i|x_i}\}_{a_i}$ employed and the fully separable state $\rho =  \sum_{\lambda} P(\lambda) \rho^{\lambda}_1 \otimes \rho^{\lambda}_2 \otimes \cdots \otimes \rho^{\lambda}_n$ shared, the resulting $\vecP$ always satisfies the above inequality. Consequently, if the observed $\vecP$ gives rise to a violation of the Bell inequality $\I_n$, one can immediately conclude that the shared state $\rho$ is entangled: this is the observation that allows one to employ a Bell inequality as a device-independent entanglement witness~\cite{Bancal11}.

\subsection{Entanglement depth, entanglement intactness, and their device-independent certification}

As mentioned above, entanglement in a multipartite scenario is much more complicated. To this end, if an $n$-partite pure state $\ket{\psi}$ can be separated into $m$ tensor factors, i.e., 
\begin{equation}\label{Eq:GenericPureState}
	\ket{\psi} = \ket{\phi^{(1)}} \otimes \ket{\phi^{(2)}} \otimes \cdots \otimes \ket{\phi^{(m)}}, 
\end{equation}
then it is said to be $m$-separable~\cite{Horodecki:RMP}.  Producibility is a closely related concept but focuses instead on the number of parties in each subgroup: $\ket{\psi}$ is said to be $k$-producible~\cite{Guhne_Produciblity_05} if the number of parties defining each tensor factor $\ket{\phi^{(i)}}$ is at most $k$-partite. For example, any $n$-partite state $\ket{\psi}$ is, according to the definition given, trivially 1-separable and $n$-producible. In contrast, a fully separable $n$-partite state is $n$-separable as well as being $1$-producible. These examples make it evident that one is generally interested in the smallest $k$ for which a given state $\ket{\psi}$ is $k$-producible but not $(k-1)$-producible: the quantity $k$ is then known as the entanglement depth~\cite{Anders_EntanglementDepth_01} of $\ket{\psi}$. Likewise, one is interested in the largest $m$ for which a given state $\ket{\psi}$ is $(m-1)$-separable but not $m$-separable: the quantity $m$ is then known as the entanglement intactness~\cite{He:PRX:2018} of $\ket{\psi}$. An $n$-partite GME state is one that has an entanglement depth of $n$ and an entanglement intactness of 1.

The separability and producibility for mixed states are similarly defined. If a density matrix $\rho$ can be written as a convex mixture of pure states that are $m$-separable (respectively $k$-producible), then we say that it is $m$-separable (respectively $k$-producible), thus, the set of $m$-separable (respectively $k$-producible) states are convex. Moreover, the set of $m$-separable ($k$-producible) states is a subset (superset) of $m'$-separable ($k'$-producible) set for all $m'\le m$ ($k' \ge k$). With some thoughts, one realizes that the convexity of these sets translates, via Eq.~\eqref{eq:BornRule}, into the convexity of the set of $\vecP$ that can be obtained from $m$-separable ($k$-producible) states (if we impose no restriction on the Hilbert space dimension). The separating hyperplane theorem then dictates that for any $\vecP$ that does not belong to the $m$-separable ($k$-producible) set for some fixed $m$ ($k$), one can construct a {\em linear} witness to separate $\vecP$ from the corresponding set. In other words, we may use a Bell inequality, Eq.~\eqref{eq:LocalCorrelations}, or, more precisely, the strength of violation of a Bell inequality to certify that a given $\vecP$ cannot arise from any $m$-separable ($k$-producible) state for some fixed $m$ ($k$), thus putting an upper bound (lower bound) on the entanglement intactness (depth) of the underlying state. Put it differently, for any Bell inequality $\I_n$, if one can determine $k$-producible bound:
\begin{equation}
  \label{eq:KproducibleBound}
	S_n^{\text{$k$-prod.}}:=\max_{\vecP'\in\text{$k$-prod. set}} \sum_{\vec{a},\vec{x}}\beta_{\vec{a}}^{\vec{x}}P'(\vec{a}|\vec{x}),
\end{equation}
then an empirical observation of $\sum_{\vec{a},\vec{x}}\beta_{\vec{a}}^{\vec{x}}P(\vec{a}|\vec{x})>S_n^{\text{$k$-prod.}}$ would certify an entanglement depth of the underlying state that is at least $k+1$. Since this conclusion holds without invoking any assumption about the dimension of the underlying state $\rho$, let alone the measurements employed, the inequality
\begin{equation}\label{Eq:DIWED:generic}
	\sum_{\vec{a},\vec{x}}\beta_{\vec{a}}^{\vec{x}}P(\vec{a}|\vec{x})\stackrel{k\text{-prod.}}{\le} S_n^{\text{$k$-prod.}}
\end{equation}
serves as a DIWED by placing a {\em lower bound} of $k+1$ on the underlying entanglement depth. In the terminology of Ref.~\cite{Curchod2015}, this means that Eq.~\eqref{Eq:DIWED:generic} is a constraint that has to be satisfied by a nonlocal quantum resource of minimal group size $k$. 

Likewise, a device-independent witness for entanglement intactness:
\begin{equation}\label{Eq:m-sep-witness}
	\sum_{\vec{a},\vec{x}}\beta_{\vec{a}}^{\vec{x}}P(\vec{a}|\vec{x})\stackrel{m\text{-sep.}}{\le} S_n^{\text{$m$-sep.}}
\end{equation}
can be established by determining the $m$-separable bound:
\begin{equation}
  \label{Eq:m-sep-bound}
	S_n^{\text{$m$-sep.}}:=\max_{\vecP'\in\text{$m$-sep. set}} \sum_{\vec{a},\vec{x}}\beta_{\vec{a}}^{\vec{x}}P'(\vec{a}|\vec{x}).
\end{equation}	
A violation of the witness given in Eq.~\eqref{Eq:m-sep-witness} then allows one to put a device-independent {\em upper bound} of $m-1$ on the entanglement intactness of the underlying state.

The concepts of $m$-separability and $k$-producibility are evidently closely related. In fact, it is easy to see that~\cite{Guhne_Produciblity_05} a quantum state that is $k$-producible is necessarily $m$-separable for $m\le\lceil\tfrac{n}{k}\rceil$ while a quantum state that is $m$-separable is necessarily $k$-producible for $k\ge\lceil\tfrac{n}{m}\rceil$. In particular, if we are to determine Eq.~\eqref{eq:KproducibleBound} for $k=n-1$, it is equivalent to computing Eq.~\eqref{Eq:m-sep-bound} for $m=2$. Similarly, if we are to determine Eq.~\eqref{eq:KproducibleBound} for $k=2$, it is equivalent to computing Eq.~\eqref{Eq:m-sep-bound} for $m=n-1$. Bearing this in mind, we remark that multipartite Bell inequalities whose $m$-separable bounds have been determined can already be imported to serve as DIWEDs. A particularly worth noting example of this is the family of Mermin-Ardehali-Belinskii-Klyshko (MABK) inequalities~\cite{Mermin_MerminIneq_90,Ardehali90,Roy-Singh92,Belinski_MerminIneq_93,GisinPLA1998}, whose $m$-separable bounds for an arbitrary value of $m<n$ have been determined (see Refs.~\cite{Nagata:PRL:2002,Yu:PRL:2003}). Turning the argument around, we see that Bell inequalities for which the $k$-producible bounds  have been determined, such as those given in Ref.~\cite{Liang:PRL:2015}, can also be used to certify an upper bound on entanglement intactness.

\subsection{Reduction for pure states}

In Ref.~\cite{Liang:PRL:2015}, it was left as an open problem whether the entanglement depth of an arbitrary $n$-partite pure state can be certified in a device-independent manner. Here, we shall demonstrate that the problem reduces to that of certifying device-independently the genuine $n$-partite entanglement for {\em all} $n$-partite GME states. To this end, let us remind that with an appropriate choice of local bases, a multipartite pure quantum state $\ket{\psi}$ can always be cast in the form of Eq.~\eqref{Eq:GenericPureState} for some choices of $m$ and $\{\ket{\phi^{(i)}}\}_{i=1}^m$ such that (1) $m$ is its entanglement intactness, and (2) the maximal size\footnote{Here, size  refers to the number of subsystems involved in the definition of $\ket{\phi^{(i)}}$.} of $\ket{\phi^{(i)}}$ is the entanglement depth of $\ket{\psi}$.

Suppose that for an arbitrary $k$-partite GME pure state $\ket{\Psi}$, there exists a Bell type inequality $\I_{\ket{\Psi}}$ whose violation can be used to certify the GME nature of $\ket{\Psi}$. Now, let $\ket{\Psi}$ be the tensor factor of $\ket{\psi}$ which determines its entanglement depth, i.e., 
\begin{equation}
	\argmax_i \text{size}(\ket{\phi^{(i)}})=\ket{\Psi}.
\end{equation}
Then, a DIWED that can be used to certify the entanglement depth of $\ket{\psi}$ is  given by $\I_{\ket{\Psi}}$, when applied to the subsystems (i.e., the partial trace) of $\ket{\psi}$ that give $\ket{\Psi}$. If necessary, one could trivially extend this $k$-partite DIWED to make it an $n$-partite DIWED by lifting~\cite{Pironio_Lifting_05} the corresponding Bell-inequality to involve the remaining $(n-k)$ parties.

To this end, let us remark that the recent work of Ref.~\cite{Zwerger_DICGME_18} has indeed provided a generic recipe for the construction of such a Bell type inequality for an arbitrary GME pure state. Unfortunately, their construction requires one to perform a Bell test that generally involves many measurement settings. Thus, there remains the problem of finding tractable DIWEDs for an arbitrary pure GME pure state. For the family of $n$-partite GHZ states, 
\begin{equation}\label{Eq:GHZ}
	\ket{\text{GHZ}_n}= \frac{1}{\sqrt{2}}\left(\ket{0}^{\otimes n} + \ket{1}^{\otimes n} \right),
\end{equation}	
the problem is solved using the family of MABK inequalities~\cite{Mermin_MerminIneq_90,Ardehali90,Roy-Singh92,Belinski_MerminIneq_93,GisinPLA1998}, or the witnesses given in Ref.~\cite{Liang:PRL:2015}. For a generic multipartite pure state, however, there is no known systematic construction that involves only a few measurement settings. In what follows, we start by exploring the family of DIWEDs given in Ref.~\cite{Liang:PRL:2015} and its one-parameter generalization to understand its usefulness when it comes to witnessing the GME nature of other families of states. Later, in Sec.~\ref{sec:DIWEDforGraph}, we provide a systematic construction that allows us to witness the GME nature of one-dimensional cluster states.


\section{A one-parameter family of DIWEDs}
\label{sec:FamilyIneq}

In Ref.~\cite{Liang:PRL:2015}, the following family of Bell expressions was proposed as the first example of a DIWED:
\begin{equation}\label{Eq:EDWitnessesGeneralized}
   \S_{n,\gamma}:=\frac{\gamma}{2^n}\left(\sum_{\vec{x}\in\{1,2\}^n} E_n(\vec{x})\right)-E_n(\vec{2}_n),\quad  0<\gamma\le 2,
\end{equation}
where $\vec{2}_n:=(2,2,\ldots, 2)$ is an $n$-bit vector of twos. While the analysis in Ref.~\cite{Liang:PRL:2015} concerned predominantly the case of $\gamma=2$, we show in Appendix~\ref{app:LocalBound} that $\S_{n,\gamma}$ has a local bound of 1 for all $\gamma\in(0,2]$. When $n=\gamma = 2$, Eq.~\eqref{Eq:EDWitnessesGeneralized} leads to the well-known Clauser-Horne-Shimony-Holt (CHSH) Bell inequality~\cite{CHSH}:
\begin{equation}
  \label{eq:CHSH}
  \frac{1}{2}\sum_{\vec{a},\vec{x}}(-1)^{(x_1-1)(x_2-1)}a_1 a_2 P(\vec{a}|\vec{x}) \overset{\L}{\leq}1.
\end{equation}

Let $\S^{\Q,*}_{n,\gamma}$ denote the maximal quantum value of $\S_{n,\gamma}$. It was shown in Ref.~\cite{Liang:PRL:2015} that for $k\le n$, the $k$-producible bound of $\S_{n,\gamma}$, [cf. Eq.~\eqref{eq:KproducibleBound}] is precisely $\S^{\Q,*}_{k,\gamma}$ and thus depends only on $k$, but not on $n$. Numerically, for $n\le 6$, we have found that the maximal quantum value of $\S_{n,\gamma}$ can always be achieved by considering the $n$-partite GHZ state $\ket{\GHZ_n}$ and the following ansatz~\cite{Werner_FullCorrelatorsIneqs_2001,Liang:PRL:2015} of measurement observables:
\begin{equation}
  \label{eq:MeasurementsOfGHZ}
  \begin{array}{rcl}
    A_{x_i = 1} &=& \cos{\alpha}\,\sigma_x + \sin{\alpha}\,\sigma_y, \\
    A_{x_i = 2} &=& \cos(\phi_n + \alpha)\,\sigma_x + \sin(\phi_n + \alpha)\,\sigma_y, \\
  \end{array}
\end{equation}
where $\alpha = -\frac{n-1}{2n}\phi_n$ and $\phi_n \in [0,\frac{\pi}{2}]$. 
The resulting quantum value, computed via $E_n(\vecx)=\bra{\GHZ_n}\otimes_{i=1}^n A_{x_i}\ket{\GHZ_n}$, then reads as:
\begin{equation}\label{Eq:SQMax}
  \S^{\Q,*}_{n,\gamma} = \gamma\,\cos^{n+1}\left(\frac{\phi^*_n}{2}\right) - \cos\left(\frac{n+1}{2}\phi^*_n\right),
\end{equation}
where the analytic form of the optimal parameter $\phi^*_n$ and optimal quantum value $\S^{\Q,*}_{n,\gamma}$ for $2\le n \le 5$ can be found in Appendix~\ref{app:ExplicitQBounds}.  Note that for these values of $n$, $\S^{\Q,*}_{n,\gamma}$ appear to increase monotonically with $n$ as well as $\gamma$, thus showing a gap between the $k$ and $k+1$ producible bounds for $k=1,2,3,4,5$. In Ref.~\cite{Liang:PRL:2015}, the presence of such a gap was numerically verified for $k \le 7$ but only for the case of $\gamma=2$. 

\subsection{Effectiveness for some families of qubit states}

Under the assumption that there always exist the aforementioned gaps between successive $k$-producible bounds and noting that the {\em maximal} quantum violation of a full correlation Bell inequality~\cite{Werner_FullCorrelatorsIneqs_2001,Bancal12}, such as the one associated with Eq.~\eqref{Eq:EDWitnessesGeneralized}, is always achievable~\cite{Werner_FullCorrelatorsIneqs_2001} using a GHZ state, we deduce that the GME nature of a GHZ state can always be certified using the DIWED 
\begin{equation}\label{Eq:Liang:DIWEDs}
	\S_{n,\gamma}\stackrel{k\text{-prod.}}{\le} \S^{\Q,*}_{k,\gamma}
\end{equation}
by setting $k$ to be $n-1$.

\begin{figure}
  \includegraphics[scale=0.17]{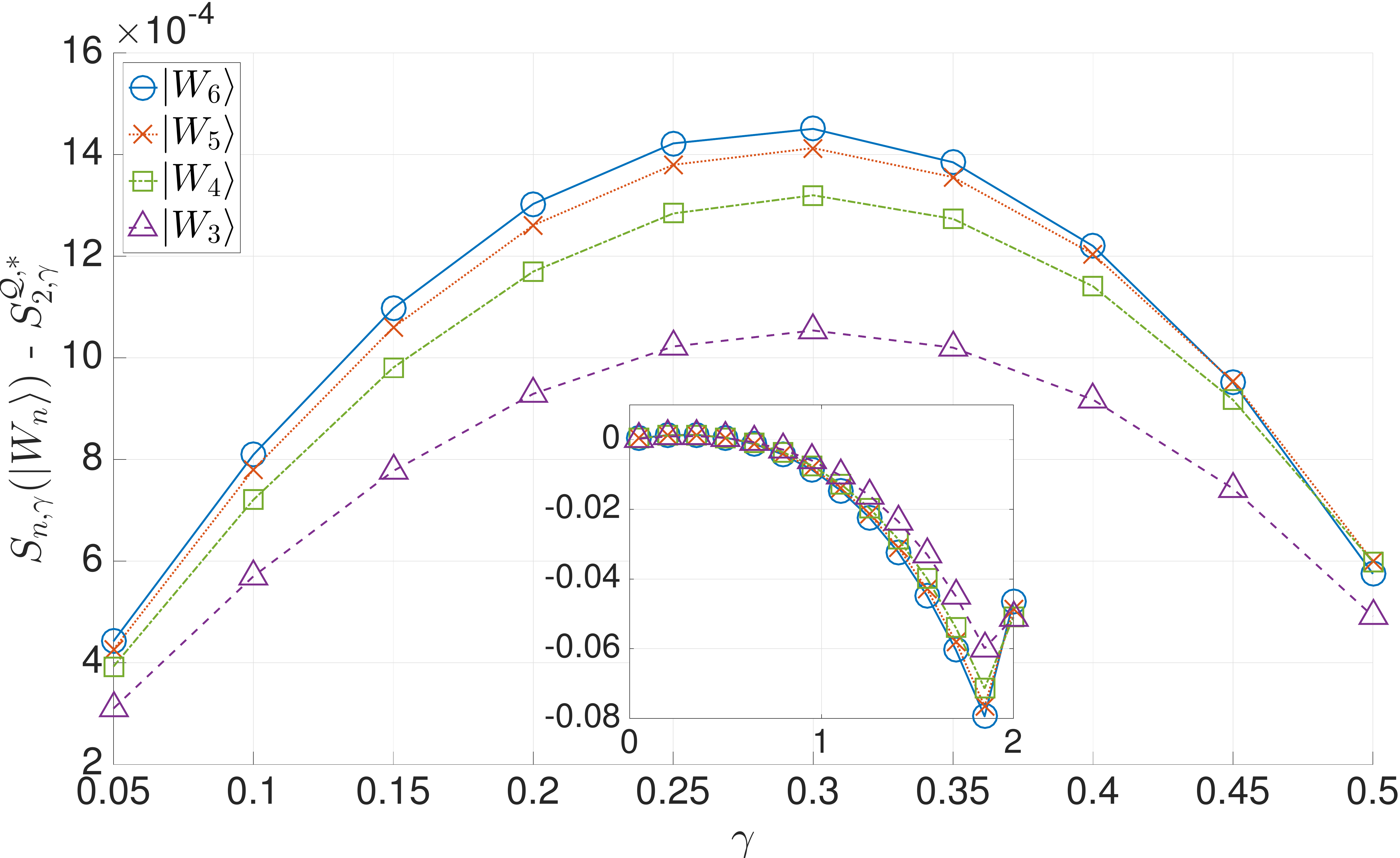}
  \caption{
Plots of the quantum violation of the DIWED of Eq.~\eqref{Eq:EDWitnessesGeneralized} achieved by $\ket{W_n}$ in the restricted interval of $\gamma\in[0.05,0.5]$ (the inset gives the same plots but for the whole interval of $\gamma\in(0,2]$ where one finds that the witnesses are generally not violated, i.e., giving a negative value).
\label{fig:TuningGamma_ED}}
\end{figure}

In Ref.~\cite{Liang:PRL:2015}, the DIWEDs of Eq.~\eqref{Eq:Liang:DIWEDs} with $\gamma=2$ were also considered in conjunction with the $W$ states $\ket{W_n}$, and the one-dimensional cluster states (i.e., graph states corresponding to a ring graph $\ket{{\rm RG}_n}$ or a linear chain  $\ket{{\rm LG_n}}$). Unfortunately, the numerical results presented therein suggested that for $3\le n\le 7$ ($3< n\le 7$), the witness can, at best, be used to certify that $\ket{W_n}$ ($\ket{{\rm RG}_n}$ and $\ket{{\rm LG_n}}$) is entangled.\footnote{Note that in Table II of Ref.~\cite{Liang:PRL:2015}, the certifiable entanglement depth of $\ket{W_3}$ was mistaken to be 3, even though the result presented in Table III  in the corresponding Supplemental Material clearly indicate that this should be 2. } In other words, the witness of Eq.~\eqref{Eq:Liang:DIWEDs} with $\gamma=2$ essentially failed to detect {\em any} of the more-than-two-body entanglement present in all these multipartite qubit states.

\begin{figure*}
  \includegraphics[width = 1\textwidth]{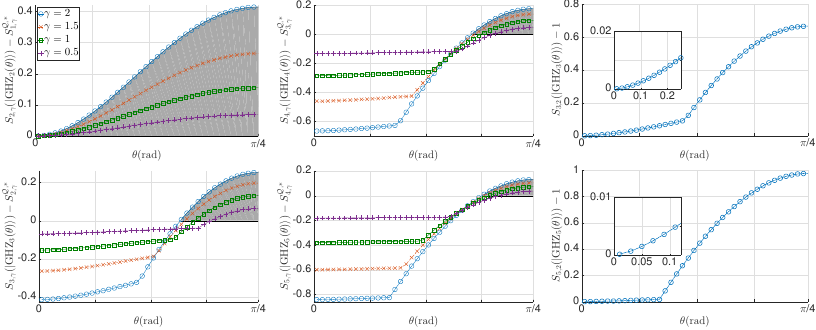}
  \caption{ \label{fig:WeightedGHZ} (Left and middle plots) Best quantum value found for $\S_{n,\gamma}(\ket{{\rm GHZ}_{n}(\theta)}) - S_{n-1,\gamma}^{\Q,*}$ for $\theta\in(0,\frac{\pi}{4}]$, $\gamma=0.5,1,1.5,2$, and $n=2,3,4,5$. As $\gamma$ decreases from 2, the GME nature of $\ket{{\rm GHZ}_n(\theta)}$ is certifiable for a more restricted range of $\theta$. We shade the region of the plots where the GME nature of the states are correctly certified. (Rightmost plots) Best quantum value found for $\S_{n,2}(\ket{{\rm GHZ}_{n}(\theta)})$ less the local bound of 1 for $n=3$ and 5. Since this difference is always non-negative, we always observe a Bell-inequality violation of $\ket{{\rm GHZ}_{n}(\theta)}$ for all $\theta\in(0,\frac{\pi}{4}]$ that we have tested, showing a strong contrast with the results of Ref.~\cite{Zukowski:PRL:2002}. In the insets, we zoom-in to region of $\theta$ where the consideration of Ref.~\cite{Zukowski:PRL:2002}, which allow only nontrivial measurements, would fail to show the violation of the corresponding entangled states.}
\end{figure*}

Would tuning the parameter $\gamma$ help in providing a better lower bound on the entanglement depth of any of these states? In the case of $\ket{W_n}$, this indeed turned out to be of some use. By numerically maximizing~\cite{YC_SeeSaw_07} the quantum value of $\S_{n,\gamma}$ for $\ket{W_n}$ over the POVMs used to define qubit measurements, we have found that for (some instances of) $\gamma\in[0.05,\,0.5]$, the optimized quantum value of $\S_{n,\gamma}$ for $\ket{W_n}$ with $n=3,4, 5,6$ indeed exceeds the corresponding 2-producible bound $\SQmax{2}$, thus correctly certifying the GME nature of $\ket{W_3}$ while improving, in comparison with the $\gamma=2$ DIWED, the entanglement depth certifiable for $\ket{W_4}$, $\ket{W_5}$ and $\ket{W_6}$. Note, however, that these violations of the witness are tiny, and should thus be seen as a proof-of-principle demonstration that tuning the free parameter can indeed be useful in some cases. The details of these findings are shown in Fig.~\ref{fig:TuningGamma_ED}.

Unfortunately, tuning the parameter does not seem to help at all for the device-independent certification of the entanglement depth of  the one-dimensional cluster states. In Sec.~\ref{sec:DIWEDforGraph}, we return to these states and present other DIWEDs that are naturally suited for them.

Now, let us focus, instead, on the following family of unbalanced-weight GHZ states:
\begin{equation}\label{Eq:GHZ:theta}
    \ket{{\rm GHZ}_{n}(\theta)} =\cos{\theta}\ket{0}^{\otimes n} + \sin{\theta}\ket{1}^{\otimes n},\quad \theta\in\left(0,\tfrac{\pi}{4}\right],
\end{equation}
where $\ket{{\rm GHZ}_{n}}$ defined in Eq.~\eqref{Eq:GHZ} corresponds to the special case of $\ket{{\rm GHZ}_{n}(\theta=\frac{\pi}{4})}$.

For $n=2$, it is known~\cite{Gisin_HoldsInequality1991} that all such states violate the CHSH Bell inequality. However, for $n\ge 3$, it is also known~\cite{Zukowski:PRL:2002} that for $n$ odd and $\sin2\theta\le 2^{\frac{1-n}{2}}$, $\ket{{\rm GHZ}_{n}(\theta)}$ {\em cannot} violate any full correlation Bell inequality with rank-1 qubit projective measurement. It thus seems unlikely to certify the entanglement depth of {\em all} these GME states using the DIWED of Eq.~\eqref{Eq:EDWitnessesGeneralized}. In fact, our numerical results, see Fig.~\ref{fig:WeightedGHZ}, suggest that the interval of $\theta$ for which we can correctly certify the GME nature of $\ket{{\rm GHZ}_{n}(\theta)}$ shrinks as $n$ increases. Again, tuning the value of the parameter $\gamma$ does not seem to help.

In contrast with the results of Ref.~\cite{Zukowski:PRL:2002}, we have nonetheless found that by allowing trivial (degenerate qubit) measurements, for (at least) $n=2,3,4,5$, it is possible to demonstrate the nonlocality of all entangled $\ket{{\rm GHZ}_{n}(\theta)}$. As an example, let us consider $\ket{{\rm GHZ}_3(\theta=0.07~{\rm rad})}$. Note that each nontrivial qubit observable $\hat{m}\cdot\vec{\sigma}$ can be parametrized by a unit vector $\hat{m}\sim (\vartheta,\varphi)$ where $\vartheta$, $\varphi$ are, respectively, the polar angle and the azimuthal angle of $\hat{m}$. To exhibit the nonlocality of the aforementioned three-partite  state, it suffices for the three parties to set their first measurement observable to be the one associated, respectively,  with the unit vectors of (0.7734 $\pi$ rad., 0.6767 $\pi$ rad.), (0.7457 $\pi$ rad., 0.1533 $\pi$ rad.), and (0.2295 $\pi$ rad., 0.8300 $\pi$ rad.) while the second observable, respectively, as $\sigma_z$, $\Id$, and $-\sigma_z$. Note that since trivial measurements are involved to violate the inequality for these states, there must exist some non-full-correlation Bell inequalities which are also violated by these states. However, such Bell inequalities (with marginal correlators) may depend on the particular state in question. Also, since the violations are tiny, this should be seen as a proof-of-principle demonstration that the nonlocality of $\ket{{\rm GHZ}_n(\theta)}$ can still be witnessed for even nearly separable $\ket{{\rm GHZ}_n(\theta)}$ by a full-correlation Bell inequality.

\subsection{Effectiveness for higher-dimensional GHZ states}
\label{sec:GeneralizedGHZ}
Next, we discuss the effectiveness of using Eq.~\eqref{Eq:EDWitnessesGeneralized} to certify the ED of the $n$-partite $d$-dimensional GHZ states $\ket{{\rm GHZ}_{n,d}}$. Importantly, each of these GME states can actually be seen as the direct sum of $n$-qubit GHZ state $\ket{{\rm GHZ}_{n}}$ residing in disjoint qubit subspaces (and a product state if $d$ is odd):
\begin{equation}
    \label{eq:DecomposeGHZ}
  \begin{split}
    \ket{{\rm GHZ}_{n,d}} = &\frac{1}{\sqrt{d}}\sum_{i=0}^{d-1}\ket{i}^{\otimes n}\\ 
    = &\sqrt{\frac{2}{d}}\left(\bigoplus_{j=0}^{\lfloor d/2-1 \rfloor} \ket{{\rm GHZ}^{(j)}_n}  \oplus \frac{\chi}{\sqrt{2}}\ket{d-1}^{\otimes n}\right),
  \end{split}
\end{equation}
where $\ket{{\rm GHZ}^{(j)}_n}$ is the $n$-qubit GHZ state acting on the local qubit subspace spanned by $\{\ket{2j},\ket{2j+1}\}$, $\chi=1$ if $d$ is odd but vanishes otherwise.

In view of this, it is  not surprising that the certifiable ED apparently depends on the parity of the local Hilbert space dimension. 

In particular, the very same maximal quantum value of Eq.~\eqref{Eq:EDWitnessesGeneralized}, i.e., $\S^{\Q,*}_{n,\gamma}$ given in Eq.~\eqref{Eq:SQMax}, is attainable using $\ket{{\rm GHZ}_{n,d}}$ whenever the local Hilbert space dimension $d$ is {\em even}. Explicitly, this can be achieved using the following choice of block-diagonal qudit observables:
\begin{align}\label{eq:EvenGHZ}
  A^{(d)}_{x_i=1}&=\bigoplus_{j=0}^{\frac{d}{2}-1} A_{x_i=1},  \quad A^{(d)}_{x_i=2}=\bigoplus_{j=0}^{\frac{d}{2}-1} A_{x_i=2},
\end{align}
where $A_{x_i=1}$, $A_{x_i=2}$ are defined in Eq.~\eqref{eq:MeasurementsOfGHZ},  the $j$-th qubit observable in each of these direct sums acts on the qubit subspace spanned by $\{\ket{2j},\ket{2j+1}\}$, and the full correlator~\cite{Werner_FullCorrelatorsIneqs_2001, Bancal12} is computed as $E_n(\vecx)=\bra{\GHZ_{n,d}}\otimes_{i=1}^n A^{(d)}_{x_i}\ket{\GHZ_{n,d}}$. Thus, the DIWED of  Eq.~\eqref{Eq:EDWitnessesGeneralized} can correctly certify the ED of such states,  as with the case for $\ket{{\rm GHZ_{n}}}$. 

On the other hand, when the local Hilbert space dimension is {\em odd}, the ED certifiable using the DIWED of Eq.~\eqref{Eq:EDWitnessesGeneralized}, based on our numerical results, is not tight. However, these bounds on ED do become tighter and approach the actual ED  as the Hilbert space dimension increases.  Specifically, with the following choice of block-diagonal qudit observables:
\begin{equation}\label{eq:oddGHZ}
\begin{split}
  A^{(d)}_{x_i=1}&=\bigoplus_{j=0}^{\frac{d}{2}-1} A_{x_i=1}\oplus 1,  \\
  A^{(d)}_{x_i=2}&=\left\{\begin{array}{ll} 
  \bigoplus_{j=0}^{\frac{d}{2}-1} A_{x_i=2}\oplus 1, \quad i\neq n,\\
  \bigoplus_{j=0}^{\frac{d}{2}-1} A_{x_i=2}\oplus -1, \quad i=n,\\
  \end{array}
  \right.
\end{split}
\end{equation}
we recover the best quantum value $\frac{d-1}{d}\S^{\Q,*}_{n,\gamma} + \frac{1}{d}$ that we have  found for these states.
 
\begin{figure}[h!]
  \includegraphics[scale = 0.35]{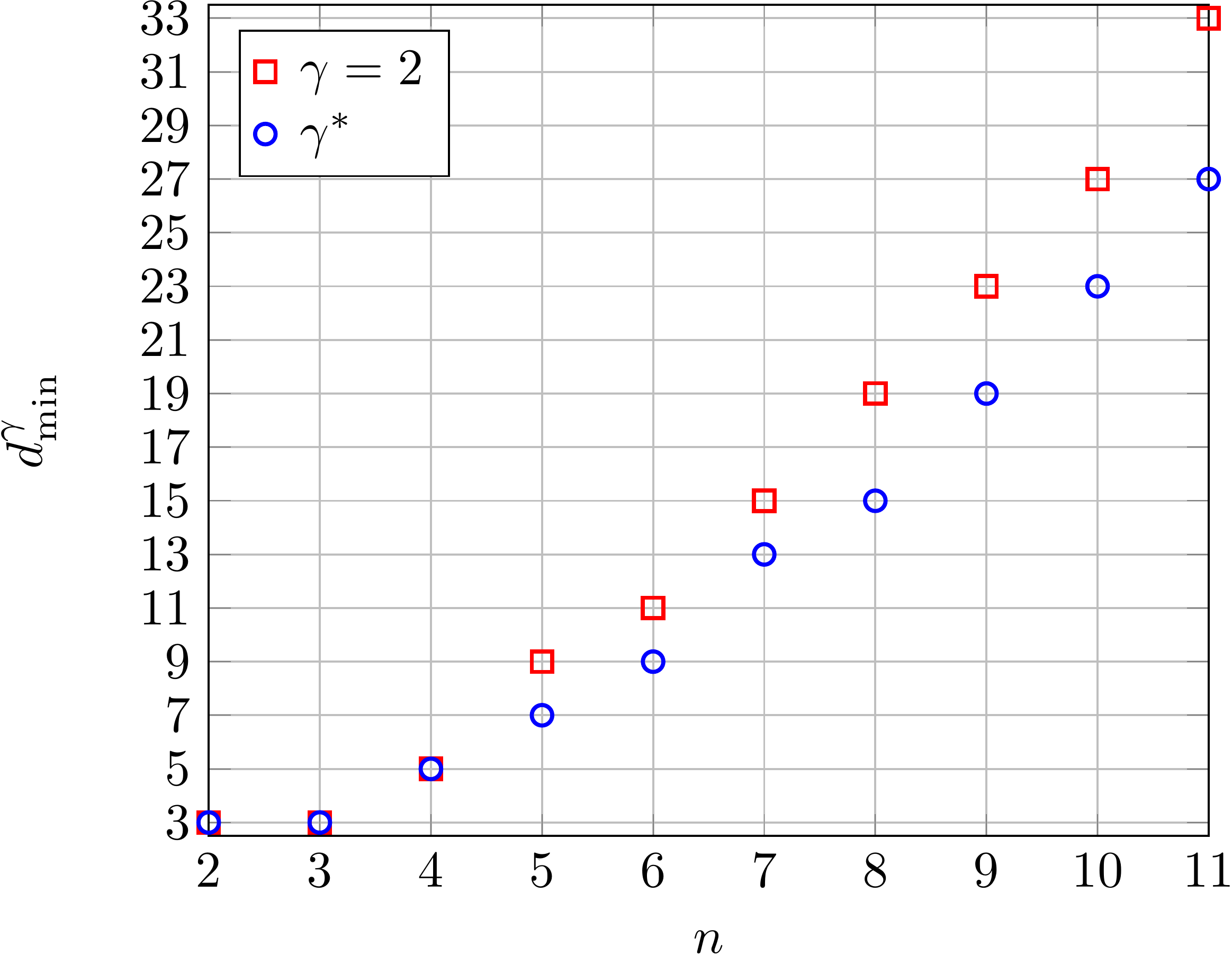}
  \caption{\label{fig:D_DimensionGHZ}Plot of the combinations of $(n,d^{\gamma = 2}_{\rm min})$ and $(n,d^{\gamma^*}_{\rm min})$  where the GME nature of $\ket{{\rm GHZ}_{n,d}}$ can be certified using the DIWED of Eq.~\eqref{Eq:EDWitnessesGeneralized} when $\gamma$ is set, respectively, to 2 and an optimized value $\gamma^*$.}
\end{figure}

The effectiveness of the DIWED of Eq.~\eqref{Eq:EDWitnessesGeneralized} in certifying the entanglement depth of $\ket{{\rm GHZ}_{n,d}}$ for $d$ odd can then be decided by verifying if the following inequality holds true for the given local Hilbert space dimension $d$ and for some $\gamma \in (0,2]$: 
\begin{equation}
  \label{eq:CriteriaforGammaD}
  (d-1)\S^{\Q,*}_{n,\gamma} - d\,\S^{\Q,*}_{n-1,\gamma} \stackrel{?}{>} -1.
\end{equation}
To this end, note that if the above equation holds for some value of $n$, $\gamma$, and $d=d^\gamma_{\rm min}$, then it must also hold for the same combination of $n$, $\gamma$, and $d>d^\gamma_{\rm min}$. 
In Fig.~\ref{fig:D_DimensionGHZ}, we plot for each $n\le 11$, the value of $d^{\gamma=2}_{\rm min}$ and the smallest value of $d^\gamma_{\rm min}$ found by optimizing the choice of $\gamma$ (with the optimal value denoted by $\gamma^*$). As with the case of $\ket{W_n}$, we have found that for $\ket{{\rm GHZ}_{n,d}}$ with $n\ge 5$, tuning the parameter $\gamma$ can sometimes lead to a tighter lower bound on the ED of these latter states.

\section{DIWED for graph states}
\label{sec:DIWEDforGraph}

\subsection{Graph states and stabilizers}

We now return to the problem of certifying, in a device-independent manner, the GME nature of graph states. Specifically, we shall demonstrate how one can construct a non trivial Bell-inequality suited for certifying the ED of certain qubit graph states. Before that, let us first recall from Ref.~\cite{Hein2004} the definition of a qubit graph state. For a given graph, with $n$ vertices and edges, consider the following tensor product of qubit operators
\begin{equation}\label{Eq:GraphStabilizers}
    g_i = X_i \underset{j \in N(i)}{\otimes} Z_j,\quad i\in \{1,\dots,n\},
\end{equation}
where $N(i)$ is the set of neighbors of vertex $i$, $X_i$, $Y_k$, and $Z_j$ represent, respectively, the Pauli matrix $\sigma_x$, $\sigma_y$, and $\sigma_z$ acting on the $i$th, $k$th, and $j$th qubit. For simplicity, we have omitted in the above equation the identity operator $\Id$ acting on the remaining (non-neighboring) qubits. The $g_i$ of Eq.~\eqref{Eq:GraphStabilizers} is known as a stabilizer of the corresponding graph state $\ketG$, which is defined as the simultaneous $+1$ eigenstate of all stabilizers $g_i$.

We should further introduce the stabilizer set $S({\rm G}) = \{s_j,j=1,\dots,2^n\}$ formed by all possible combinations of the product of $g_i$:
\begin{equation}
  s_j = \prod_{i\in V_j({\rm G})}g_i,
\end{equation}
where $V_j({\rm G})$ is a subset of vertices of the graph G. Notice that since $g_i$ commutes with $g_{i'}$ for all $i,\; i'$,  the order of the products does not matter. The corresponding $\ketG$ is evidently also the $+1$ eigenstate of all $s_j\in S({\rm G})$.

\subsection{Nonlocality of graph states}
\label{Sec:3input}

The reason of introducing $S({\rm G})$ is that we can use its elements to construct a Bell operator~\cite{Braunstein1992}, which can be translated into a Bell inequality naturally suited for $\ketG$.
The idea was first proposed in~\cite{Guhne_IneqForGraphStates_05}, where they summed over all elements of $S({\rm G})$ to obtain the Bell operator:
\begin{equation}
  \BFull{{G}} = \sum_{i=1}^{2^n} s_i.
  \label{eq:BFull}
\end{equation}

\begin{figure}[t]
  \includegraphics[scale=0.35]{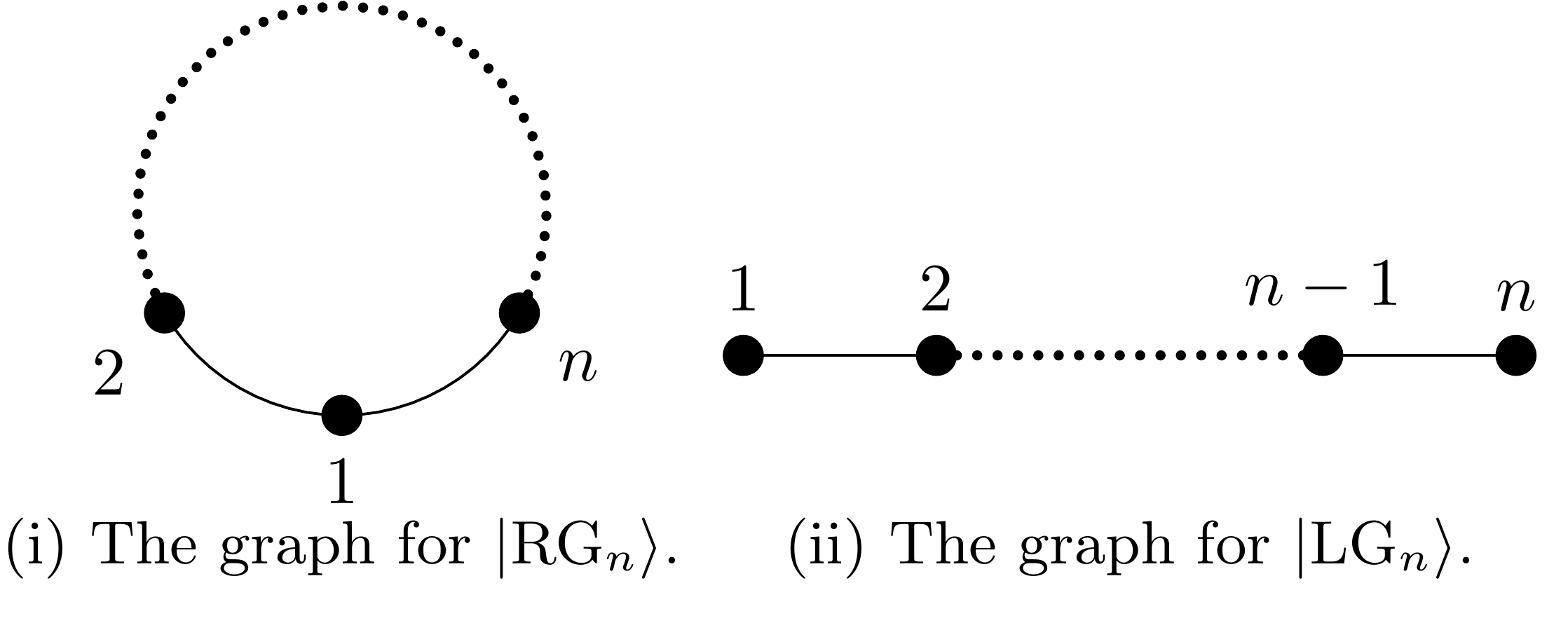}
\caption{\label{fig:PicturesofGraph} Two families of $n$-node graphs: (i) the ring graph gives rise to a graph state that is equivalent to a one-dimensional cluster state with a periodic boundary condition, (ii) the linear graph gives rise to a graph state that is equivalent to a one-dimensional cluster state with a closed boundary condition. For $n=3$, it is known~\cite{Hein2004} that these two kinds of graph states are both local-unitarily (LU) equivalent to $\ket{{\rm GHZ}_3}$. Likewise, for $n=4$, if we allow also the symmetry of \emph{graph isomorphisms} (i.e., the permutation of parties), these two kinds of graph states are again LU equivalent~\cite{Hein2004} to each other, but not to $\ket{{\rm GHZ}_4}$.}
\end{figure}

To get a Bell expression in the form of Eq.~\eqref{eq:LocalCorrelations}, one associates, to each term $s_i$, the Pauli matrices $X,Y, Z$, respectively, with measurement setting $1,2,3$. Each term $s_i$ is then mapped to a correlator $E_k(\vec{x})$ with $k\le n$. Importantly, depending on the graph G, the corresponding correlator of $s_i$ may be a {\em marginal} correlator, i.e., one involves only nontrivial measurements on a strict subset of subsystems and hence $k<n$. Graphs of $n$ nodes arranged in a ring (straight line), giving rise to the $n$-partite ring (linear) graph states $\ket{{\rm RG}_n}$ ($\ket{{\rm LG}_n}$), are some explicit examples of this kind (see Fig.~\ref{fig:PicturesofGraph}). By determining the local bound of the corresponding Bell expression, one then obtains a Bell inequality whose maximal quantum value of $2^n$ is guaranteed to be achievable by the underlying graph state $\ketG$~\cite{Guhne_IneqForGraphStates_05}.

\subsection{Two-setting DIWED for graph states based on GHZ paradox}
\label{sec:WitnessTwo}

Evidently, from an experimental points of view, it would be desirable to make use of a Bell inequality with fewer measurement settings (or fewer expectation values to be measured).  To this end, we  follow the proposal of Ref.~\cite{Scarani_NonlocalityCluster_05} to construct Bell inequalities by choosing only a subset of the elements in $S({\rm G})$, specifically those that allow us to demonstrate the so-called GHZ paradox~\cite{Kafatos1989-KAFBTQ, Mermin_90}. Let us stress that as with the work of Ref.~\cite{Guhne_IneqForGraphStates_05}, Ref.~\cite{Scarani_NonlocalityCluster_05} only concerns the demonstration of nonlocality for graph states, but here we wish to go a step further by using such a Bell inequality as a DIWED to witness the ED of the corresponding $\ketG$.

Consider, without loss of generality,  a situation where $m$ distinct correlators take on their maximal value, 
\begin{equation}\label{Eq:PerfectCorrelation}
	E_{n}(\vec{x}^{'})=E_{n}(\vec{x}^{''})=...=E_{n}(\vec{x}^{''\cdots'})=+1.
\end{equation}
For simplicity, in Eq.~\eqref{Eq:PerfectCorrelation} and in the following arguments, we write all the correlators as $n$-partite correlators, but this is only for the convenience of presentation, rather than a necessary requirement of the arguments. In particular, some of these correlators could well be marginal correlators that involve less than $n$ parties. A GHZ paradox, also known as a proof of nonlocality without inequality, arises {\em if} Eq.~\eqref{Eq:PerfectCorrelation} implies also $E_{n}(\vec{y})=1$ for Bell-local correlations but $E_{n}(\vec{y})=-1$ for certain quantum correlation. 

To appreciate when such a ``paradox" may occur, let us consider extremal\footnote{For the case of nonextremal $\vecP\in\L$ satisfying Eq.~\eqref{Eq:PerfectCorrelation}, one first decomposes $\vecP$ into all extremal $\vecP$ of $\L$ satisfying Eq.~\eqref{Eq:PerfectCorrelation} and repeats the following arguments.} points of $\L$, i.e., those where the outcome $a_{x_j}$ of the $x_j$th measurement of the $j$th party is deterministic, taking either $\pm1$ for all $j$, and all $x_j$. Then, it follows from Eq.~\eqref{Dfn:Correlator} that each equation in Eq.~\eqref{Eq:PerfectCorrelation}, such as,
\begin{equation}\label{Eq:DeterministicDecomposition}
	E_n(\vec{x}^{'})=a_{x_1'}a_{x_2'}\cdots a_{x_n'}=1,
\end{equation}
imposes nontrivial constraints on the value of $a_{x_j'}$. In particular, if $\vec{y}$ is chosen such that 
\begin{equation}\label{Eq:xtoy}
	a_{y_1}a_{y_2}\cdots a_{y_n} = \prod_{i_1} a_{x_{i_1}'} \prod_{i_2} a_{x_{i_2}''}\cdots \prod_{i_m} a_{x_{i_m}^{''\cdots'}},
\end{equation}
then by virtue of Eqs.~\eqref{Eq:PerfectCorrelation} and~\eqref{Eq:xtoy}, we must also have $E_{n}(\vec{y})=1$. In other words, if $\vec{y}$ is chosen as the string of inputs where each $x_j$ only appears an odd number of times in Eq.~\eqref{Eq:PerfectCorrelation}, then these conditions of perfect correlations guarantee also $E_{n}(\vec{y})=1$.

To complete the argument of the paradox, one must find quantum correlation satisfying Eq.~\eqref{Eq:PerfectCorrelation} and giving $E_{n}(\vec{y})=-1$. While the constraints of Eq.~\eqref{Eq:PerfectCorrelation} are easily enforced by choosing the observables associated with each correlator from a stabilizer of the graph state of interest, the latter constraint of perfect {\em anti}correlation (and the requirement of having only two measurements per party) can only be achieved with a careful selection of the stabilizers corresponding to Eq.~\eqref{Eq:PerfectCorrelation}. For any such selection, a nontrivial Bell inequality  violated quantum mechanically up to the algebraic maximum of $m+1$ can then be constructed as:
\begin{equation}
  \label{Eq:Inequality_GHZ_Paradox}
	E_{n}(\vec{x}^{'})+E_{n}(\vec{x}^{''})+...+E_{n}(\vec{x}^{''\cdots'})-E_{n}(\vec{y})\stackrel{\L}{\le}m-1\stackrel{\Q}{\le}m+1.
\end{equation}
A proof of the local bound is given in Appendix~\ref{App:LocalBound:GHZParadox}, whereas the quantum bound (which is also the algebraic maximum) follows from our {\em assumption} that a GHZ paradox can be demonstrated quantum mechanically, i.e., one can find a quantum strategy that makes the first $m$ correlator take value +1 and the last correlator take value -1. In what follows, we provide a construction of such Bell inequalities for various graph states where each of the first $m$ correlators is identified with an element of $S({\rm G})$ while the last term is identified with the product of all these chosen $s_i$'s.

\subsubsection{Ring graphs}

As a first example, let us consider $\ket{{\rm RG}_3}$. From its stabilizer set $S({\rm RG}_3)$, we choose the three stabilizers $g_1$, $g_2$, $g_3$ and their product $g_1g_2g_3$ to define our Bell operator, i.e.,
\begin{equation}
	\B_{{\rm RG}_3}=\sum_{i=1}^3 g_i + \prod_{i=1}^3 g_i.
\end{equation}
Writing these operators explicitly while associating $X$ with the first measurement and $Z$ with the second measurement gives:
\begin{equation}
  \begin{split}
    \label{eq:ExampleGHZParadox}
    &g_1 = X_1Z_2Z_3 \Rightarrow E_3(1,2,2), \\
    &g_2 = Z_1X_2Z_3 \Rightarrow E_3(2,1,2), \\
    &g_3 = Z_1Z_2X_3 \Rightarrow E_3(2,2,1), \\
    &g_1g_2g_3 = -X_1X_2X_3 \Rightarrow -E_3(1,1,1).
  \end{split}
\end{equation}
Taking the sign of each term into account, we obtain
\begin{equation}
	\IPart{{RG}_3}:=E_3(1,2,2) + E_3(2,1,2) + E_3(2,2,1) - E_3(1,1,1),
\end{equation} 
which is exactly a representative of the MABK Bell expression~\cite{Mermin_MerminIneq_90, Ardehali90, Roy-Singh92, Belinski_MerminIneq_93, GisinPLA1998}, whose quantum 2-producible bound~\cite{Curchod2015} is known~\cite{Nagata:PRL:2002} to be $2\sqrt{2}$. Note that when this bound is saturated, it may still be possible to certify genuine tripartite entanglement in a device-independent manner if marginal distributions are taken into account~\cite{Bhattacharya2017}.

The resulting Bell inequality thus has the following properties:
\begin{equation}
  \label{eq:BellIneqRG3_Part_Complete}
    \IPart{{RG}_3}\overset{\L}{\leq}2 \stackrel{2\text{-prod.}}{\le}2\sqrt{2}\,\, \overset{\Q}{\leq}4,
\end{equation}
thus making $\IPart{{RG}_3}\stackrel{2\text{-prod.}}{\le} 2\sqrt{2}$ a DIWED that can be used to certify the GME nature of $\ket{{\rm RG}_3}$.

More generally, we propose to consider the following Bell operator, each involving $n$ stabilizers from $S({\rm RG}_n)$ and their product:
\begin{equation}
  \label{eq:IneqForRingGraph}
  \begin{split}
    \BPart{{\rm RG}_{n = {\rm odd}}}&= \sum_{i=1}^n g_i + \prod_{i=1}^{n}g_i; \\
    \BPart{{\rm RG}_{ n = {\rm even}}}&= g_n\left(1+\sum_{i=1}^{n-1} g_i\right) +\prod_{i=1}^{n-1}g_i. 
  \end{split}
\end{equation}
In Appendix~\ref{app:FormulaForRingGraph}, we show that each of these Bell operators only involves two different qubit measurements per party\footnote{Note that, if any of the parties performs only one measurement during the Bell experiment, as we show in Appendix~\ref{app:OneMeasurementBISep}, such a Bell expression can never serve as a DIWED for the corresponding graph state as the 2-separable bound always coincides with the Tsirelson bound.}. For the case of odd $n$, these are always $X$ and $Z$. The same observation applies for party 2, 3, $\dots$, $n-2$ in the case of even $n$, but for  party 1, $n-1$, and $n$, these become, respectively, $\{Y, Z\}$, $\{Y, Z\}$ and $\{X,Y\}$.

Throughout, we shall adopt the following convention in mapping a stabilizer to the corresponding correlator: if in the Bell operator (and hence the list of stabilizers) considered, only $X_j$ and $Y_j$ are involved, we associate these operators, respectively, as the first and second measurement of the $j$th party; if, instead, only $X_j$ and $Z_j$ are involved, we associate them, respectively, as its first and second measurement; and if only $Y_j$ and $Z_j$ are involved, we associate them, respectively, as its first and second measurement.

Thus, for odd $n$, the resulting Bell expression reads as:
\begin{equation}\label{Eq:Bell:RG:odd}
	\IPart{{RG}_n}:=E_n(2,1,2,\varnothing,\ldots,\varnothing) + \circlearrowright - E_n(\vec{1}_n),
\end{equation}
where $E_n(2,1,2,\varnothing,\ldots,\varnothing):=E_3(2,1,2)$ is a triparite correlator involving only the first three parties\footnote{Here and below, we use the symbol $\varnothing$ to indicate the {\em trivial} measurement setting.} and $\circlearrowright$ is a short hand to denote the additional $n-1$ terms that need to be included to make the Bell expression invariant under arbitrary {\em cyclic} permutation of parties (cf. Ref.~\cite{Grandjean2012}). 

In a similar manner, we note from $\BPart{{\rm RG}_{n =4}}$ that it only involves $Y_1, Z_1, X_2, Z_2, Y_3, Z_3, X_4, Y_4$ (see Appendix~\ref{app:FormulaForRingGraph} for details). We thus arrive at the following Bell expression for $n=4$:
\begin{align}\label{Eq:Bell:RG:n=4}
    \IPart{{RG}_4}:=&E_4(2,\varnothing,2,1) + E_4(\varnothing,1,\varnothing,1) + E_4(2,2,1,2)\nonumber\\
	 + &E_4(1,2,2,2) - E_4(1,1,1,\varnothing).
\end{align}
For $n=6$, $\BPart{{\rm RG}_{n =6}}$ involves only $Y_1, Z_1, Y_5, Z_5, X_6, Y_6$ and $X_i, Z_i$ for $i=2,3,4$. The corresponding Bell expression is (see Appendix~\ref{app:FormulaForRingGraph} for details):
\begin{align}\label{Eq:Bell:RG:n=6}
      \IPart{{RG}_6}&:=E_6(2,\varnothing,\varnothing,\varnothing,2,1)+ E_6(1,2,\varnothing,\varnothing,2,2) \nonumber\\
      & + E_6(\varnothing,1,2,\varnothing,2,1) +  E_6(2,2,1,2,2,1)\\
      & +  E_6(2,\varnothing,2,1,\varnothing,1) +  E_6(2,\varnothing,\varnothing,2,1,2) - E_6(\vec{1}_5,\varnothing).\nonumber
\end{align}

Using the numerical technique of Ref.~\cite{Moroder13}, upper bound on the $k$-producible bounds of $\IPart{{RG}_n}$ for all $k<n\le 6$ can be computed. The results are summarized in Table~\ref{tab:KProducibleBounds_RG_LG}. In particular, it is worth noting that the (upper bound of the) $(n-1)$-producible bound always appear to be  $n-(3-2\sqrt{2})<n+1$, thus showing that the constructed Bell inequality can indeed serve as a DIWED for $\ket{{\rm RG}_n}$.

\subsubsection{Fully-connected graphs}

Apart from $\ket{{\rm RG}_n}$, the symmetry presented in fully-connected, i.e., complete graphs also allows us to derive a simple yet nontrivial family of Bell inequalities for the corresponding graph states. For  concreteness, let us denote the state of an $n$-partite fully connected graph  by $\ket{{\rm FG}_n}$.  Although $\ket{{\rm FG}_n}$ is known~\cite{Hein2004} to be LU equivalent to $\ket{{\rm GHZ}_n}$, which we already know how to certify its entanglement depth using the DIWED of  Sec.~\ref{sec:FamilyIneq}, the simplicity of this family of Bell expressions warrants a separate discussion. To this end, consider the Bell operator
\begin{equation}
  \label{eq:BellOperators_Fullgraph}
    \BPart{{\rm FG}_n} = \sum_{i=1}^{n}g_i+ g_1g_2g_3, 
\end{equation}
which is easily verified to involve both $X$ and $Z$ measurements for all parties.  We thus have the following Bell expression:
\begin{equation}
  \label{Ineq:Fullgraph}
    \IPart{{\rm FG}_{n}} = E_n(1,2,2,\dots,2) + \circlearrowright' - E_n(\vec{1}_3,\vec{2}_{n-3}),
\end{equation}
where we have used $\circlearrowright'$ to indicate the additional $n-1$ terms that need to be included to ensure that the first $n$ terms are invariant under arbitrary {\em cyclic} permutation of parties.
Note that when $n=3$, $\IPart{{\rm FG}_{n=3}}$ is again a special case of the MABK inequalities~\cite{Mermin_MerminIneq_90, Belinski_MerminIneq_93}.

A remark is now in order. In our previous construction of Bell inequalities based on the GHZ paradox, the value of $E_n(\vec{y})$ is determined by all other terms appearing in the Bell expression. However, this is not the case for $\IPart{FG_{n}}$. In fact, the GHZ paradox can already be exhibited using $E_n(1,\vec{2}_{n-1})$, $E_n(2,1,\vec{2}_{n-2})$, $E_n(\vec{2}_2,1,\vec{2}_{n-3})$ and $E_n(\vec{1}_3,\vec{2}_{n-3})$ alone.  The remaining terms are included to ensure that all parties perform two alternative measurements. Despite this difference, the local bound and quantum bound are easily shown, respectively, to be $n-1$ and $n+1$ (see Appendix~\ref{App:LocalBound:GHZParadox} for a proof of the local bound).

\subsubsection{Other types of graphs}
\label{Sec:OtherGraphs}

Beyond the two families of graph states presented above, appropriate DIWED can also be constructed to certify (at least for $n$ up to 6) the entanglement depth of $\ket{{\rm LG}_n}$, i.e., the graph state associated with a linear chain (see Fig.~\ref{fig:PicturesofGraph}). Since $\ket{{\rm LG}_3}$ is LU equivalent to $\ket{\GHZ_3}$, its entanglement depth can already be certified using the DIWED discussed in Sec.~\ref{sec:FamilyIneq}. Likewise, we omit the discussion of $\ket{{\rm LG}_4}$, since it is LU equivalent to $\ket{{\rm RG}_4}$ if we allow also the permutation of parties.
On the other hand, our construction leads to the following Bell expressions, respectively,  for $n=5$ and 6: 
\begin{equation}\label{Eq:BellExp:LG5}
  \begin{split}
    \IPart{{\rm LG}_5}:= & E_5(\varnothing,2,1,2,\varnothing) +  E_5(\varnothing,\varnothing,\varnothing,2,1)\phantom{+E()} \\
    +& E_5(1,2,\varnothing,2,1) + E_5(1,\varnothing,2,1,2)\\
    + &E_5(2,1,1,1,2)- E(2,1,2,2,\varnothing), 
    \end{split}
  \end{equation}
\begin{equation}
  \begin{split}
    \IPart{{\rm LG}_6}:=&  E_6(\varnothing,\varnothing,2,1,2,\varnothing)+ E_6(\varnothing,\varnothing,\varnothing,2,1,2) \\
    + & E_6(\varnothing,\varnothing,\varnothing,\varnothing,2,1)+ E_6(2,1,2,2,1,2)\\
    + & E_6(2,2,1,\varnothing,1,2) + E_6(1,2,\varnothing,1,\varnothing,1)\\
    - & E_6(1,1,1,\varnothing,1,2).  
  \end{split}
\end{equation}  
In Appendix~\ref{app:IneqsLinearGraphs}, we provide the Bell operators and the explicit form of the individual stabilizers leading to these Bell expressions.
The $k$-producible bounds for these Bell expressions are summarized in Table~\ref{tab:KProducibleBounds_RG_LG}. One should note that, in general, the subset of $S({\rm G})$ chosen to construct a DIWED via our procedure is not unique. In particular, we provide in Appendix~\ref{app:IneqsLinearGraphs} an alternative (inequivalent) Bell expression that is also suitable for the certification of the entanglement depth of $\ket{{\rm LG}_5}$.  In this regard, note also that some of these choices may lead to a Bell expression where one (or more) of the parties only has one fixed measurement setting. In Appendix~\ref{app:OneMeasurementBISep}, we show that such Bell expressions are generally not useful for a device-independent certification of the entanglement depth of the corresponding graph state, as its Tsirelson bound, assuming some mild condition holds, can also be saturated using a biseparable state.

\begin{figure}[t]
\includegraphics[scale =0.35]{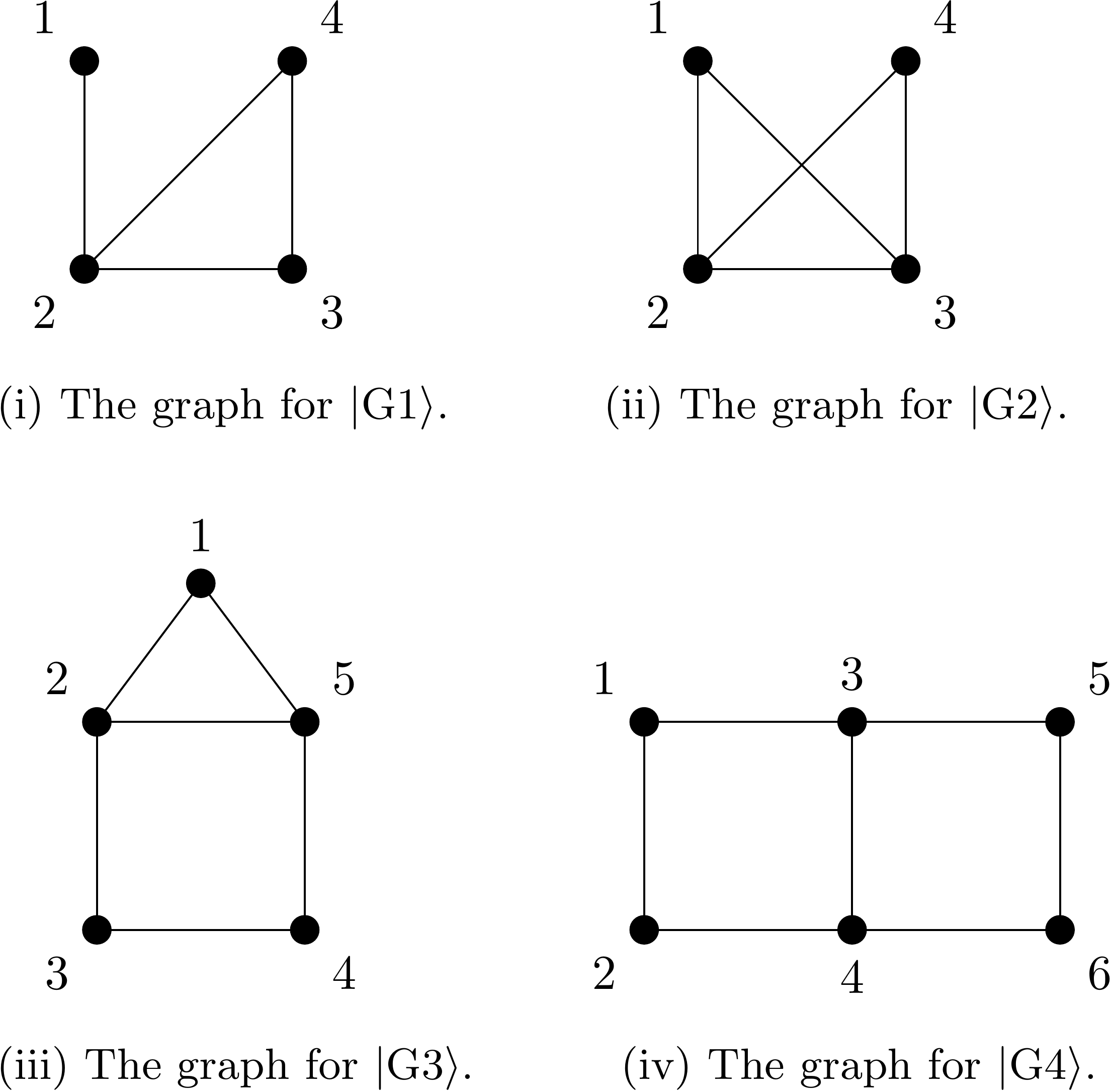}
  \caption{ \label{fig:OtherGraphs} The other graph states that have been considered in this work. 
For these asymmetrical graphs, the labels of the nodes are especially crucial as they are used to define the stabilizers [cf. Eq.~\eqref{Eq:GraphStabilizers}] involved in the corresponding Bell operator.
}
\end{figure}
\begin{table}[h!]
\begin{center}
  \begin{tabular}{|c|c|c|c|c|c|} 
    \hline
    & 2-prod. & 3-prod. & 4-prod. & 5-prod. & 6-prod.\\ [0.5ex] 
    \hline\hline
    $\IPart{RG_3}$&  $^\ddag2\sqrt{2}$ & *4 & - & - & -\\ 
    \hline
    $\IPart{{\rm RG_4}}$ &  $^\ddag2\sqrt{2}+1$ & $^\ddag2\sqrt{2}+1$ & *5 & - & - \\
    \hline
    $\IPart{{\rm RG_5}}$ & $^\ddag$4 & 4.0855 & $^\ddag2\sqrt{2}+2$ & *6 & - \\
    \hline
    $\IPart{{\rm RG_6}}$ & $^\ddag$5 & 5.0453 & 5.0660 & $2\sqrt{2}+3$ & *7 \\
    \hline \hline
    $\IPart{{\rm LG_5}}$&  4.0602 & $^\ddag2\sqrt{2}$ + 2  & $^\ddag2\sqrt{2}$ + 2 & *6 & -\\ 
    \hline
    $\IPart{{\rm LG_6}}$& $^\ddag2\sqrt{2}$+3 & $^\ddag2\sqrt{2}$ + 3  & $^\ddag2\sqrt{2}$ + 3 &  $^\ddag2\sqrt{2}$ + 3 & *7\\ 
    \hline\hline
   
    $\IPart{{\rm FG_4}}$& $^\ddag$3.6742 &  $^\ddag$4.4037 & *5 & - &-\\
    \hline
    $\IPart{{\rm FG_5}}$& $^\ddag$4.6188 & $^\ddag$5.1962 & $^\ddag5.4037$ & *6 & -\\ 
    \hline
    $\IPart{{\rm FG_6}}$& $^\ddag$5.5902 & $^\ddag$6.0977  & $^\ddag$6.1962 & $^\ddag$6.4037 &*7\\
    \hline
\end{tabular}
\end{center}
\caption{\label{tab:KProducibleBounds_RG_LG} Summary of \emph{upper} bounds on the quantum $k$-producible bounds of $\IPart{RG_n}$, $\IPart{LG_n}$ and $\IPart{FG_n}$ for $n$ up to 6.  Here, and in the following tables, entries marked with $^\ddag$ correspond to upper bounds that have been numerically verified to be {\em tight}, i.e., achievable (within the numerical precision of the solver) using $k$-producible quantum states. By construction, the Tsirelson bounds (marked with $*$, i.e., the $n$-producible bounds) are always tight. Apart from the 2-producible bound of $\IPart{RG_3}$ and all the Tsirelson bounds, all analytic expressions presented are extracted from the results of our numerical optimization.}
\end{table}

For completeness, we have also  considered the graph states associated with the  three remaining 4-node (connected) graphs, [see Figs.~\ref{fig:OtherGraphs}(i), ~\ref{fig:OtherGraphs}(ii), and the star graph,  as well as a 5-node graph [Fig.~\ref{fig:OtherGraphs}(iii)], and a 6-node graph [Fig.~\ref{fig:OtherGraphs}(iv)]]. The graph state associated with the 4-node star graph state is known~\cite{Hein2004} to be LU equivalent to $\ket{{\rm GHZ}_4}$, whose entanglement depth is certifiable using the DIWED of Sec.~\ref{sec:FamilyIneq}, we thus omit it from the following discussion. Likewise, the graph state associated with G1, and G2, are both known~\cite{Hein2004} to be LU equivalent to $\ket{\text{LG}_4}$ (and hence to $\ket{\text{RG}_4}$ if we also allow the permutation of parties), a DIWED constructed for $\ket{\text{RG}_4}$ can thus also be used to witness the entanglement depth of $\ket{\text{G1}}$ and $\ket{\text{G2}}$.

The Bell expressions that can be used as DIWEDs for the remaining two states are, respectively,
\begin{equation}
  \begin{split}
    \IPart{G3}:= & E_5(\varnothing,\varnothing,2,1,2)+  E_5(1,\varnothing,\varnothing,2,1) \phantom{+E(as\,\,)}\\
    + &E_5(2,2,1,2,2) + E_5(2,1,\varnothing,1,\varnothing) \\
    + &E_5(\varnothing,2,2,2,1)  - E_5(1,1,1,2,\varnothing),
  \end{split}
\end{equation}
and
  \begin{equation}
    \begin{split}
      \IPart{G4}:=& E_6(2,1,\varnothing,2,\varnothing,\varnothing)+ E_6(\varnothing,2,2,1,\varnothing,2)   \\
      + &  E_6(\varnothing,\varnothing,\varnothing,2,2,1) + E_6(1,2,1,2,2,\varnothing) \\
      + & E_6(2,\varnothing,1,2,1,2)  - E_6(1,1,2,1,1,1).
    \end{split}
  \end{equation}
The quantum $k$-producible bounds for these Bell expressions are summarized in Table~\ref{tab:KProducibleBounds_OG_FG}. In Appendix~\ref{app:OtherGraphs}, one can find the Bell operators and the explicit form of the stabilizers leading to these Bell expressions.
  
    \begin{table}[h!]
\begin{center}
  \begin{tabular}{|c|c|c|c|c|c|c|} 
    \hline
    & 1-prod. & 2-prod. & 3-prod. & 4-prod. & 5-prod. & 6-prod.\\ [0.5ex]
\hline
    $\IPart{{\rm G3}}$ & 4 & $\ddag4$ & 4.2330 & $^\ddag2\sqrt{2}+2$ & *6 &- \\
    \hline
    $\IPart{{\rm G4}}$ & 4 & 4.1225 & 4.2833 & $2\sqrt{2}+2$ & $^\ddag2\sqrt{2}+2$ & *6 \\
    \hline
  \end{tabular}
\end{center}
\caption{\label{tab:KProducibleBounds_OG_FG} Summary of {\em upper} bounds on the quantum $k$-producible bounds of $\IPart{G3}$ and $\IPart{G4}$. The graph associated with $\IPart{Gi}$ can be found in Figure~\ref{fig:OtherGraphs}.}
\end{table}

\section{Conclusions and Outlook}
\label{sec:Conclusions}

In this paper, we have explored various two-setting, two-outcome Bell inequalities for the device-independent certification of the entanglement depth of several (families of) pure genuine multipartite entangled (GME) states. To this end, we first revisited the one-parameter family of Bell expressions introduced in Ref.~\cite{Liang:PRL:2015} and turned them into device-independent witnesses for entanglement depth (DIWED) by determining, for six or less parties,  the quantum $k$-producible bounds of these Bell expressions. 

We then showed that, in comparison with the DIWEDs originally introduced in Ref.~\cite{Liang:PRL:2015}, these parametrized DIWEDs can, via an appropriate choice of the parameter $\gamma$, witness tighter entangled depth of the $W$ states $\ket{W_n}$. They can also be  used to witness the entanglement depth of certain unbalanced-weight GHZ states $\ket{{\rm GHZ}_n(\theta)}$, even though tuning the value of $\gamma$ does not seem to help in this case. It is particularly worth noting that, with the help of trivial measurements, the full-correlation Bell inequalities associated with these witnesses are apparently violated by {\em all} entangled $\ket{{\rm GHZ}_n(\theta)}$ (for $n\le 5$), thereby suggesting a way to get around the no-go result presented in Ref.~\cite{Zukowski:PRL:2002}.

Beyond qubits, we have found that these witnesses can perfectly witness the entanglement depth of higher-dimensional generalization of the GHZ states, namely, 
$\ket{{\rm GHZ}_{n, d}}$, whenever the local Hilbert space dimension $d$ is even. For odd $d$, these witnesses do not seem to always witness a tight entanglement depth, but their power grows with $d$; again, tuning the parameter $\gamma$ improves the ability of these DIWEDs to certify the right entanglement depth. 

Next, we turned our attention to graph states. By using the properties of the graph states and the idea of a GHZ paradox~\cite{Kafatos1989-KAFBTQ, Mermin_90}, we constructed Bell inequalities that are maximally violated by the corresponding graph states. For ring graphs and the fully connected graphs (i.e., complete graphs) of $n$ nodes, we provided a general construction and showed that the resulting DIWEDs can indeed be used to certify the GME nature of the corresponding graph states for $n\le 6$. We conjecture that the same conclusion holds for $n>6$. Similarly, we have also provided DIWEDs to witness the GME nature of graph states for linear graphs (up to 6 nodes) and a few other graphs with a small number of nodes.

A few possibilities for future research stem naturally from this work. First, while our numerical results have shown that a very weakly entangled generalized GHZ state  (i.e., $\ket{{\rm GHZ}_n(\theta)}$ with $\theta\to 0$) can violate a full correlation Bell inequality for $n=2, 3, 4$ and 5, it would be interesting to show that this holds  indeed for an arbitrary number of parties (thereby completely lifting the no-go result of Ref.~\cite{Zukowski:PRL:2002} for full correlation Bell inequalities). On a related note, it is also worth showing that for witnessing the entanglement depth of $\ket{{\rm GHZ}_n(\theta)}$ using the one-parameter family of witnesses,  $\gamma=2$ indeed represents the optimal choice of parameter. For $W$ states $\ket{W_n}$, it may be worth combining the numerical techniques of Ref.~\cite{Moroder13} and the results of Ref.~\cite{Brunner12} to find simple, yet better DIWED that can be used to certify the entanglement depth of these states for arbitrary $n\ge 4$. 

Evidently, it would be desirable to determine the tight, quantum $(n-1)$-producible bounds for the family of $n$-partite Bell expressions that we have constructed for the ring graph states as well as that for the fully connected graph states. This would allow one to address the conjecture that there is always a gap between these bounds and the quantum maximum of $n+1$. In the same spirit, finding a simple, systematic construction of DIWED for the linear graph states would be more than welcome. 

Admittedly, a drawback of the construction that we have presented for graph states is there is always only a constant gap of two between the local bound and the quantum maximum. This means that as $n$ increases, the white-noise robustness of these DIWEDs would decrease with $n$. To cope with that, one could consider, instead, three-setting DIWEDs based on the construction discussed in Sec.~\ref{Sec:3input}. Some explicit examples of this can be found in Appendix~\ref{app:WitnessThreeSettings}. Which (among these two) kinds of DIWEDs would be more favorable in an actual experimental situation? Answering this clearly requires a separate investigation taking into account various experimental constraints, a problem that is already outside the scope of the present work.

{\em Note added.} Recently, we became aware of the work of Ref.~\cite{Baccari2018} which also discussed the construction of two-setting Bell inequalities that are maximally violated by given graph states. 

\begin{acknowledgements}
We thank Antonio Ac\'{\i}n, Jean-Daniel Bancal, J{\k{e}}drzej Kaniewski, Denis Rosset, and Valerio Scarani for useful discussions. 
This work is supported by the Ministry of Science and Technology, Taiwan (Grants No. 104-2112-M-006-021-MY3, No. 107-2112-M-006-005-MY2, and No. 107-2627-E-006-001).

\end{acknowledgements}

\appendix

\section{Local bound}

 Before giving the proofs of local bounds, let us remind that $\L$ is a convex polytope and that the Bell expressions considered  here are linear in $E_n(\vec{x})$ (and thus in $\vecP$). Therefore a maximization of these Bell expressions over $\vecP\in\L$ can always be attained by considering {\em only} the extreme points of $\L$, i.e., local deterministic probability distributions satisfying $E_n(\vec{x})= \prod_{i=1}^{n}E_1(x_i)$ and  $E_1(x_i)=\pm1$ for all $i$ and all $x_i$.

\subsection{For the Bell expressions given in Eq.~\eqref{Eq:EDWitnessesGeneralized}}

\label{app:LocalBound}

We now give a proof that the maximal value of 
\begin{equation}
   \S_{n,\gamma}:=\frac{\gamma}{2^n} \left( \sum_{\vec{x}\in\{1,2\}^n} E_n(\vec{x})\right)-E_n(\vec{2}_n),\quad  0<\gamma\le 2,\tag{\ref{Eq:EDWitnessesGeneralized}}
\end{equation}
over the set of Bell-local correlations $\L$ is 1.  For local deterministic distributions, the first term of $\S_{n,\gamma}$ simplifies to
\begin{equation}
  \label{eq:BoundsOnFirstPart}
    \frac{\gamma}{2^n}\sum_{\vec{x}}E_n(\vec{x}) =  \frac{\gamma}{2^n}\sum_{\vec{x}}\prod_{i=1}^{n}E_1(x_i) =\frac{\gamma}{2^n}\prod_{i=1}^{n}\left[\sum_{x_i}E_1(x_i)\right].
\end{equation}
There are now two distinct cases to consider. Firstly, if $E_1(x_i=1) = - E_1(x_i=2)$ for any $i=1,2,\ldots, n$, then the above expression vanishes. $\S_{n,\gamma}$ is then maximized by having an odd number of parties setting $E_1(x_i=2)=-1$, thereby giving the value 1 as its maximum.

On the other hand, if $E_1(x_i=1) = E_1(x_i=2)$ for all $i$, then Eq.~\eqref{eq:BoundsOnFirstPart} becomes $\pm\gamma$. There are now two subcases to consider, either an even or an odd number of parties set $E_1(x_i=2)=-1$ [while the rest set $E_1(x_i=2)=1$]. In the even case, $\S_{n,\gamma}=\gamma-E_n(\vec{2}_n)= \gamma-\prod_{i=1}^{n}E_1(x_i=2)=\gamma-1\le1$ since $\gamma\le 2$. For the odd case, we have $\S_{n,\gamma}=-\gamma+1<1$ since $\gamma>0$. All in all, we thus see that $\S_{n,\gamma}$ is upper bounded by 1 for $\vecP\in\L$ whenever $\gamma\in(0,2]$ and that this bound is attainable.

\subsection{For two-outcome Bell inequalities based on GHZ paradox}
\label{App:LocalBound:GHZParadox}

We now give a proof of the following Bell inequality:
\begin{equation}
	E_{n}(\vec{x}^{'})+E_{n}(\vec{x}^{''})+...+E_{n}(\vec{x}^{''\cdots'})-E_{n}(\vec{y})\stackrel{\L}{\le}m-1.\tag{\ref{Eq:Inequality_GHZ_Paradox}}
\end{equation}
Again, it suffices to consider local extremal strategies for this purpose. By construction, 
\begin{equation}
	E_{n}(\vec{y})=E_{n}(\vec{x}^{'})E_{n}(\vec{x}^{''})\cdots E_{n}(\vec{x}^{''\cdots'}).
\end{equation}
If Eq.~\eqref{Eq:PerfectCorrelation} holds, it follows that $E_{n}(\vec{y})=1$, thus showing that the value of $m-1$ is attainable by Bell-local correlations. 
Suppose instead that $\ell$ of the $m$ of these correlators take value $-1$ then $E_n(\vec{y})$ equals to $(-1)^{\ell}$. The left-hand side of Eq.~\eqref{Eq:Inequality_GHZ_Paradox} is thus $m-2\ell - (-1)^\ell<m-1$ since $\ell>1$ by assumption. This concludes the proof of the Bell inequality of Eq.~\eqref{Eq:Inequality_GHZ_Paradox}.

The local bound of the Bell expressions $\IPart{FG_n}$, 
\begin{equation}
    \IPart{{\rm FG}_{n}} = E_n(1,2,2,\dots,2) + \circlearrowright' - E_n(\vec{1}_3,\vec{2}_{n-3}) \tag{\ref{Ineq:Fullgraph}},
\end{equation}
can be shown in a similar way. First, let us consider the case where the first $n$ terms of $\IPart{FG_n}$ is $1$. The constraints of $E_n(1,2,2,\dots,2)=E_n(2,1,2,\dots,2)=E_n(2,2,1,2,\dots,2)=1$ imply that $E_n(\vec{1}_3,\vec{2}_{n-3})=1$ and one obtains $n-1$ for $\IPart{FG_n}$. Now, if $\ell$ among the first 3 terms  and $l$ from the remaining $n-3$ terms are $-1$, then one gets $n-2\ell -2l - (-1)^\ell<n-1$ for $\IPart{FG_n}$, thus showing that $n-1$ is the local bound of $\IPart{FG_n}$.

\section{Quantum bounds of the Bell expressions given in Eq.~\eqref{Eq:EDWitnessesGeneralized} }

\label{app:ExplicitQBounds}

Here, we provide the analytic expression of the optimal parameter $\phi_n^*$ alluded to right after Eq.~\eqref{Eq:SQMax}, and the corresponding maximal quantum value $\S^{\Q,*}_{n,\gamma}$ for $n=2,\;3,\;4$, and 5. While the analytic expression for some of the larger values of $n$ can also be computed, e.g., using \emph{Mathematica}, they are too complicated to be included here. 
With formulas listed in Table~\ref{table:ExplicitQ}, one can easily verify the difference between $\S^{\Q,*}_{k,\gamma}$ and $\S^{\Q,*}_{k+1,\gamma}$ for $k=2,3,4$.

\begin{widetext}
\begin{center}
\begin{table}[h!]
\begin{tabular}{|c|c|c|}\hline
      $n$   & $\phi^*_n$(rad) & $\S^{\Q,*}_{n,\gamma}$\\
      \hline
      2   &    $2\arctan\left(-\sqrt{3-\gamma}\right)$  &    $\frac{2}{\sqrt{4-\gamma}}$\\
      3   &    $\arctan\left(\frac{4\sqrt{4-\gamma}}{\gamma}\right)$ & $\frac{8+\gamma}{8-\gamma}$ \\
      4 & $2\arctan(\frac{-\sqrt{10-\gamma-\sqrt{20+\gamma}}}{\sqrt{6+\sqrt{20+\gamma}}})$ & $\gamma(\frac{1}{6-\sqrt{20+\gamma}})^{\frac{5}{2}} - \cos(5\arctan\frac{-\sqrt{10-\gamma-\sqrt{20+\gamma}}}{\sqrt{6+\sqrt{20+\gamma}}})$ \\
      5 & $2\arctan(\frac{\sqrt{16-\gamma-\sqrt{64+6\gamma}}}{-\sqrt{16+\sqrt{64+6\gamma}}})$ & $\frac{128\sqrt{64+6\gamma}+\gamma(224+\gamma+12\sqrt{64+6\gamma})}{(-32+\gamma)^2}$\\ \hline
\end{tabular}
  \caption{\label{table:ExplicitQ}
    The analytic expression of $\phi_n^*$ and $\S^{\Q,*}_{n,\gamma}$ for $n=2,3,4,5$.
  }  
\end{table}
\end{center}

\section{Construction of $\IPart{RG_n}$}
\label{app:FormulaForRingGraph}

To see that the inequalities obtained from Eq.~\eqref{eq:IneqForRingGraph} require only two measurements per party, we provide here the explicit form of all the stabilizers involved and the corresponding correlator $E_{n}(\vec{x})$. Specifically, for odd $n$, we have
\begin{equation}
    \label{eq:StabilizersOfIRGodd}
  \begin{array}{cccccccccl}
    g_1&= &X_1&Z_2&I_3I_4\dots I_{n-2}&I_{n-1}&Z_n & \Rightarrow& E_{n}(1,2,\varnothing,\dots,\varnothing,2),\\
    g_2&= &Z_1&X_2&Z_3I_4\dots I_{n-2}&I_{n-1}&I_n & \Rightarrow& E_{n}(2,1,2,\varnothing,\dots,\varnothing),\\ 
       &&&&&&\vdots\\
    g_{n-1}&= &I_1&I_2&I_3I_4\dots Z_{n-2}&X_{n-1}&Z_n &\Rightarrow&E_{n}(\varnothing,\dots,\varnothing,2,1,2),\\
    g_{n}&= &Z_1&I_2&I_3I_4\dots I_{n-2}&Z_{n-1}&X_n & \Rightarrow& E_{n}(2,\varnothing,\dots,\varnothing,2,1),\\
    \prod_{i=1}^{n}g_i&  = & (-1)^{n-2}X_1&X_2&\dots& X_{n-1}&X_{n} &\Rightarrow& -E_{n}(\vec{1}_n),
  \end{array}
\end{equation}
where we have followed the convention of mapping a stabilizer to the corresponding correlator mentioned in the paragraph after Eq.~\eqref{eq:IneqForRingGraph}.
Note that only $X$ and $Z$ are involved in all the stabilizers of Eq.~\eqref{eq:StabilizersOfIRGodd}.

For $n=4$, we have:
  \begin{equation}
    \label{eq:StabilizersOfIRG4}
    \begin{array}{rclccc}
      g_4 &= Z_1&\Id_2&Z_3&X_4&\Rightarrow E_4(2,\varnothing,2,1), \\
      g_1g_4&=Y_1&Z_2&Z_3&Y_4& \Rightarrow E_4(1,2,2,2), \\
      g_2g_4&=\Id_1&X_2&\Id_3&X_4&\Rightarrow E_4(\varnothing,1,\varnothing,1),\\
      g_3g_4&=Z_1&Z_2&Y_3&Y_4&\Rightarrow E_4(2,2,1,2),\\
      \prod_{i=1}^3g_i&=-Y_1&X_2&Y_3&\Id_4&\Rightarrow -E_4(1,1,1,\varnothing),\\
     \end{array}
   \end{equation}
whereas for $n=6$, we have:
  \begin{equation}
    \label{eq:StabilizersOfIRG6}
    \begin{array}{rclcccccc}
      g_6 &=Z_1&\Id_2&\Id_3&\Id_4&Z_5&X_6&\Rightarrow E_6(2,\varnothing,\varnothing,\varnothing,2,1),\\      
      g_1g_6 &=Y_1&Z_2&\Id_3&\Id_4&Z_5&Y_6&\Rightarrow E_6(1,2,\varnothing,\varnothing,2,2),\\ 
      g_2g_6 &=\Id_1&X_2&Z_3&\Id_4&Z_5&X_6&\Rightarrow E_6(\varnothing,1,2,\varnothing,2,1),\\
      g_3g_6&=Z_1&Z_2&X_3&Z_4&Z_5&X_6&\Rightarrow E_6(2,2,1,2,2,1),\\
      g_4g_6&=Z_1&\Id_2&Z_3&X_4&\Id_5&X_6&\Rightarrow E_6(2,\varnothing,2,1,\varnothing,1),\\
      g_5g_6&=Z_1&\Id_2&\Id_3&Z_4&Y_5&Y_6&\Rightarrow E_6(2,\varnothing,\varnothing,2,1,2),\\
     \prod_{i=1}^5g_i&=-Y_1&X_2&X_3&X_4&Y_5&\Id_6&\Rightarrow -E_6(1,1,1,1,1,\varnothing).\\
    \end{array}
  \end{equation}
\end{widetext}
For even $n>6$, it can be deduced from the above example that only $Y_1$, $Z_1$, $X_j$, $Z_j$ for $j=2,\ldots, n-2$, $Y_{n-1}$, $Z_{n-1}$, $X_n$, and $Y_n$ are involved in the sum of the Bell operator. The resulting Bell expression is thus again one that involves only two measurement settings per party.

\section{Constructions of $\IPart{LG_n}$}
\label{app:IneqsLinearGraphs}

Here, we provide the explicit form of the stabilizers leading to the various $\IPart{LG_n}$ presented in Sec.~\ref{Sec:OtherGraphs}. Throughout, the corresponding Bell operator is always the sum of all the stabilizers listed.

For $n=5$, we have 
  \begin{equation}
    \label{eq:StabilizersOfILG5}
    \begin{array}{rclcccc}
      g_3 &=\Id_1&Z_2&X_3&Z_4&\Id_5, \\
      g_5 &=\Id_1&\Id_2&\Id_3&Z_4&X_5,\\
      g_1g_5&=X_1&Z_2&\Id_3&Z_4&X_5,\\
      g_1g_3g_4&=X_1&\Id_2&Y_3&Y_4&Z_5,\\
      g_1g_2g_3g_4&=Y_1&X_2&X_3&Y_4&Z_5,\\
      g_1g_2g_3&=-Y_1&X_2&Y_3&Z_4&\Id_5.\\
    \end{array}
  \end{equation}
Thus, only $X_1$, $Y_1$, $X_2$, $Z_2$, $X_3$, $Y_3$, $Y_4$, $Z_4$, $X_5$, and $Z_5$ are involved in $\BPart{{\rm LG}_5}$.
As mentioned in Sec.~\ref{Sec:OtherGraphs}, there could be more than one subset of $S({\rm G})$ that would achieve our goal. An example of this for $\ket{{\rm LG}_5}$ is given by:
 \begin{subequations}
\begin{equation}
	\mathcal{B}'_{{\rm LG}_5}=g_1 + g_3 + g_5 +  g_1g_3g_4 + g_2g_3g_5 +  g_2g_3g_4,
\end{equation}	
where
  \begin{equation}
    \label{eq:StabilizersOfIRG}
    \begin{array}{rclcccc}
      g_1 &=X_1&Z_2&\Id_3&\Id_4&\Id_5, \\
      g_3 &=\Id_1&Z_2&X_3&Z_4&\Id_5, \\
      g_5 &=\Id_1&\Id_2&\Id_3&Z_4&X_5,\\
      g_1g_3g_4&=X_1&\Id_2&Y_3&Y_4&Z_5,\\
      g_2g_3g_5&=Z_1&Y_2&Y_3&\Id_4&X_5, \\                 
      g_2g_3g_4&=-Z_1&Y_2&X_3&Y_4&Z_5.
    \end{array}
  \end{equation}
Thus, only $X_1$, $Z_1$, $Y_2$, $Z_2$, $X_3$, $Y_3$, $Y_4$, $Z_4$, $X_5$, and $Z_5$ are involved in $\mathcal{B}'_{{\rm LG}_5}$. Using the default convention, we arrive at
the following Bell expression:
\begin{equation}
  \begin{split}
    \I'_{{\rm LG}_5}:= &E_5(1,2,\varnothing,\varnothing,\varnothing)+ E_5(\varnothing,2,1,2,\varnothing) \\
    &+ E_5(\varnothing,\varnothing,\varnothing,2,1)+E_5(1,\varnothing,2,1,2)\\
    &+ E_5(2,1,2,\varnothing,1) - E_5(2,1,1,1,2),
    \end{split}
  \end{equation}
\end{subequations}  
which gives a Bell inequality inequivalent to Eq.~\eqref{Eq:BellExp:LG5}. The $k$-producible bounds of $\I'_{{\rm LG}_5}$ are listed in Table~\ref{Tab:LG5Prime}.
    \begin{table}[h!]
\begin{center}
  \begin{tabular}{|c|c|c|c|c|c|c|} 
    \hline
    & 1-prod. & 2-prod. & 3-prod. & 4-prod. & 5-prod.\\ [0.5ex]
    \hline
    $\I'_{{\rm LG}_5}$ &  4& $^\ddag4$ &  $2\sqrt{2}$+2 & $2\sqrt{2}$+2 & *6 \\
    \hline
  \end{tabular}
\end{center}
\caption{\label{Tab:LG5Prime}Summary of the $k$-producible bounds of $\I'_{{\rm LG}_5}$. Entries marked with $^\ddag$ have been numerically verified to be achievable with $k$-producible quantum states. By construction, the Tsirelson bounds (marked with $*$, i.e., the $n$-producible bounds) are always tight. Note that, apart from the presented 1-producible bound and the Tsirelson bound, the analytic expressions are extracted from our numerical optimization.}
\end{table}

Finally, for $n=6$, we have  
  \begin{equation}
    \label{eq:StabilizersOfILG6}
    \begin{array}{rclccccc}
      g_4 &=\Id_1&\Id_2&Z_3&X_4&Z_5&\Id_6,\\ 
      g_5 &=\Id_1&\Id_2&\Id_3&Z_4&X_5&Z_6,\\
      g_6 &=\Id_1&\Id_2&\Id_3&\Id_4&Z_5&X_6,\\
      g_2g_5&=Z_1&X_2&Z_3&Z_4&X_5&Z_6,\\
      g_2g_3g_5&=Z_1&Y_2&Y_3&\Id_4&X_5&Z_6,\\
      g_1g_2g_4g_6&=Y_1&Y_2&\Id_3&X_4&\Id_5&X_6,\\
     g_1g_2g_3g_5&=-Y_1&X_2&Y_3&\Id_4&X_5&Z_6.\\
    \end{array}
  \end{equation}
Thus, only $Y_1$, $Z_1$, $X_2$, $Y_2$, $Y_3$, $Z_3$, $X_4$, $Z_4$, $X_5$, $Z_5$, $X_6$, and $Z_6$ are involved in $\BPart{{\rm LG}_6}$.

\section{Bell operators for graph states with up to six parties}
\label{app:OtherGraphs}

Here, we provide the explicit form of the stabilizers leading to the various $\IPart{Gi}$ presented in  Sec.~\ref{Sec:OtherGraphs}. As above, the corresponding Bell operator is always the sum of all the stabilizers listed.

For $\BPart{{\rm G3}}$, we have:
  \begin{equation}
    \label{eq:StabilizersOfG3}
    \begin{array}{rclcccc}
      g_4 &=\Id_1&\Id_2&Z_3&X_4&Z_5,\\
      g_1g_5 &=Y_1&\Id_2&\Id_3&Z_4&Y_5, \\
      g_2g_3 &=Z_1&Y_2&Y_3&Z_4&Z_5,\\
      g_2g_4&=Z_1&X_2&\Id_3&X_4&\Id_5,\\
      g_2g_5&=\Id_1&Y_2&Z_3&Z_4&Y_5, \\                 
      g_1g_2g_3&=-Y_1&X_2&Y_3&Z_4&\Id_5.\\
    \end{array}
  \end{equation}
Thus, only $Y_1$, $Z_1$, $X_2$, $Y_2$, $Y_3$, $Z_3$, $X_4$, $Z_4$, $Y_5$, and $Z_5$ are involved in $\BPart{{\rm G3}}$.

  For $\BPart{{\rm G4}}$, we have:
  \begin{equation}
    \label{eq:StabilizersOfG4}
    \begin{array}{rclccccc}
      g_2 &=Z_1&X_2&\Id_3&Z_4&\Id_5&\Id_6,\\ 
      g_4 &=\Id_1&Z_2&Z_3&X_4&\Id_5&Z_6,\\ 
      g_6 &=\Id_1&\Id_2&\Id_3&Z_4&Z_5&X_6,\\
      g_1g_3&=Y_1&Z_2&Y_3&Z_4&Z_5&\Id_6,\\
      g_3g_5&=Z_1&\Id_2&Y_3&Z_4&Y_5&Z_6,\\
      g_1g_2g_4g_5g_6&=-Y_1&X_2&Z_3&X_4&Y_5&X_6.\\
    \end{array}
  \end{equation}
Thus, only $Y_1$, $Z_1$, $X_2$, $Z_2$, $Y_3$, $Z_3$, $X_4$, $Z_4$, $Y_5$, $Z_5$, $X_6$, and $Z_6$  are involved in $\BPart{{\rm G4}}$.
\section{ Uselessness of Bell expressions with only one input for some of the parties}
\label{app:OneMeasurementBISep}

In our construction, the Bell operator which naturally suits the corresponding graph state is constructed by choosing a subset from the stabilizer set such that a GHZ paradox can be shown. Among these choices, some of the resulting Bell operators require only one measurement setting for some of the parties. For example, if $g_n$ in Eq.~\eqref{eq:BellOperators_Fullgraph} for $n\ge 4$ is removed, the corresponding Bell operator gives exactly a Bell expression of this kind. When $n=4$, this leads to the following Bell operator:
\begin{equation}
  X_1Z_2Z_3Z_4 + Z_1X_2Z_3Z_4 + Z_1Z_2X_3Z_4 - X_1X_2X_3Z_4,
\end{equation}
where the last party only needs to perform $Z$ measurement while the remaining parties need to perform two measurements $\{X,Z\}$. In this case, even though the Bell expression can still witness the nonlocality of the graph state, it will generally not serve as a good DIWED to certify the genuine $n$-partite entanglement present in this state. As we show below, assuming some mild condition holds, the 2-separable bound for such Bell expressions always coincides with the Tsirelson bound, i.e., one can always find a 2-separable quantum state $\rho$ reproducing the same maximal quantum violation given by $\ketG$, thus showing that the corresponding DIWED fails to put a tight lower bound on the ED of the corresponding graph state.   

For concreteness, without loss of generality, consider the case where a Bell expression constructed from our procedure for the graph state $\ket{{\rm G}}$ is such that the last party  only performs one measurement. To exhibit the claimed GHZ paradox, this fixed measurement $\M$ is identified as one of the Pauli observables $\{X,\; Y, \;Z\}$. Let us denote the +1-eigenvalue eigenstate  of $\M$ by $\ket{+}_{\M}$ and the $(n-1)$-partite (unnormalized) state obtained by successfully projecting the last subsystem to $\ket{+}_{\M}$ by 
\begin{equation}
	\rho_{n=\ket{+}_{\M}}=\tr_n\left[\proj{{\rm G}} \left(\otimes_{j=1}^{n-1}\Id_{j}\otimes \proj{+}_{\M}\right)\right],
\end{equation}
where $\tr_n$ denotes the partial trace over the $n$th party. Henceforth, we are going to make the mild assumption that the probability of successfully performing the above projection is nonzero, i.e., $\rho_{n=\ket{+}_{\M}}\neq 0$.
Under this assumption, we will see that the normalized biseparable state:
\begin{equation}
  \label{eq:2seprho}
  \rho:= \frac{\rho_{n=\ket{+}_{\M}}\otimes \proj{+}_\M}{\tr \rho_{n=\ket{+}_{\M}}},
\end{equation}
reproduces the same correlation as that given by $\ketG$ in establishing the GHZ paradox, thus making the corresponding DIWED useless for identifying the right entanglement depth of $\ket{{\rm G}}$. 

There are now two cases to consider: one in which the stabilizer $s_i$ considered involves a Pauli observable for the last party and the other being that the stabilizer $s_i$ considered acts trivially on the last party.
We may combine the two cases by writing $s_i=\otimes_{j=1}^{n-1}M_j \otimes \M$ with  $\M, M_j\in\{X,Y,Z,\Id\}$, then
\begin{equation*}
  \begin{split}
  \tr(\rho s_i)  =& \frac{\tr\left[ \rho_{n=\ket{+}_{\M}}\otimes \proj{+}_\M (\otimes_{j=1}^{n-1}M_j \otimes \M)\right]}{\tr \rho_{n=\ket{+}_{\M}}}\\
    =& \frac{\tr\left[ \rho_{n=\ket{+}_{\M}} \left(\otimes_{j=1}^{n-1}M_j\right) \right]}{\tr \rho_{n=\ket{+}_{\M}}}\\
   =& \frac{\tr\left[ \proj{{\rm G}} \left(\otimes_{j=1}^{n-1}\Id_{j}\otimes \proj{+}_{\M}\right) \left(\otimes_{j=1}^{n-1}M_j\otimes \Id\right) \right]}{\tr \rho_{n=\ket{+}_{\M}}}\\
    =& \frac{\tr\left[ s_i\proj{{\rm G}} \left(\otimes_{j=1}^{n-1}\Id_{j}\otimes \proj{+}_{\M}\right)  \right]}{\tr \rho_{n=\ket{+}_{\M}}}\\
    =& \frac{\tr\left[ \proj{{\rm G}} \left(\otimes_{j=1}^{n-1}\Id_{j}\otimes \proj{+}_{\M}\right)  \right]}{\tr \rho_{n=\ket{+}_{\M}}}=1,\nonumber
     \end{split}
\end{equation*}

where the second equality follows by tracing out the last Hilbert space, the third equality follows by using the definition of $\rho_{n=\ket{+}_{\M}}$ and expressing the trace in the $n$-qubit Hilbert space, the fourth equality follows by invoking the cyclic property of trace and noting that $\M^2=\Id$, $ \M \proj{+}_\M  = \proj{+}_\M$, the fifth equality is a consequence of $\ket{{\rm G}}$ being an eigenvector of $s_i$ with eigenvalue 1, and the last equality follows from the definition of $\rho_{n=\ket{+}_{\M}}$. Since the Bell operator is the sum of all chosen $s_i$, and with the properties of $\rho$ shown above, the 2-separable bound attained by $\rho$ must coincide with the Tsirelson bound.

\section{Three-setting DIWED}
\label{app:WitnessThreeSettings}

We give here an illustration of how one can obtain $\IFull{G}$ from Eq.~\eqref{eq:BFull}. We use the tripartite ring graph state $\ket{{\rm RG}_3}$ to provide an explicit example, for the other states considered, we merely provide the corresponding Bell expression. 

As mentioned in Sec.~\ref{Sec:3input}, the construction of the Bell operator involves {\em all} the $2^n$ stabilizers of an $n$-partite graph state. For $\ket{{\rm RG}_3}$, the elements of $S({\rm RG}_3)$ are:
\begin{equation}
  \begin{split}
    &s_1 = X_1Z_2Z_3 = g_1, \\
    &s_2 = Z_1X_2Z_3 = g_2,\\
    &s_3 = Z_1Z_2X_3 = g_3, \\
    &s_4 = -X_1X_2X_3 = g_1g_2g_3,\\
    &s_5 = Y_1Y_2\Id_3 = g_1g_2, \\
    &s_6 = Y_1\Id_2Y_3 = g_1g_3,\\
    &s_7 = \Id_1Y_2Y_3 = g_2g_3,  \\
    &s_8 = \Id_1\Id_2\Id_3.  \\
  \end{split}
  \label{eq:ListStabilizersRG3}
\end{equation}
By associating the Pauli matrices $X_i,Y_i$, $Z_i$ and identity operator $\Id_i$, respectively, with the 1st, 2nd, 3rd measurement and $\varnothing$ of each party, each term in Eq.~\eqref{eq:ListStabilizersRG3} is mapped to a correlator $E_3(a,b,c)$, thus giving the Bell inequality $\IFull{RG3}$:        
\begin{equation}
  \label{eq:IneqRG3}
  \begin{split}
    \IFull{RG_3} := &E_3(3,1,3) -E_3(\vec{1}_3) + E_3(2,2,\varnothing) +1+ \circlearrowright\,\, \overset{\L}{\le} 6,
  \end{split}
\end{equation}
where $\circlearrowright$ is used throughout this appendix to denote the additional (4)  terms that have to be included to make the Bell expression invariant under arbitrary {\em cyclic} permutation of parties, and the constant term $E_3(\varnothing,\varnothing,\varnothing)=1$ has been left as part of the Bell expression for clarity. The local bound (equivalently, the 1-producible bound) of the Bell expression is extracted from Ref.~\cite{Guhne_IneqForGraphStates_05}. Note that when mapping a stabilizer to a correlator, as with the two-setting DIWEDs, we take the sign of each term into account when we go from Eq.~\eqref{eq:ListStabilizersRG3} to Eq.~\eqref{eq:IneqRG3}.   

Similar constructions can be obtained for $\ket{{\rm RG}_n}$ and $\ket{{\rm LG}_n}$ with $n$ up to 5.
For $\IFull{RG_4}$, we have:
\begin{align}
    \IFull{RG_4} &:= E_4(3,1,3,\varnothing) + E_4(3,2,2,3) + E_4(1,\varnothing,1,\varnothing)\nonumber\\
    &-E_4(2,1,2,\varnothing) + E_4(\vec{1}_4) +1+\circlearrowright \,\, \overset{\L}{\le} 12,
\end{align}
where $\circlearrowright$ represents another 10 terms to be included to ensure the cyclic symmetry of the Bell expression.

For $\IFull{RG_5}$, we have:
\begin{align}
    \IFull{RG_5} := &E_5(3,1,3,\varnothing,\varnothing)+ E_5(3,2,2,3,\varnothing)+E_5(1,\varnothing,1,3,3)\nonumber\\
    -&E_5(2,1,2,3,3)+E_5(2,2,\varnothing,1,\varnothing)+E_5(2,1,1,2,\varnothing)\nonumber\\
    -&E_5(\vec{1}_5) +1+\circlearrowright \,\, \overset{\L}{\le} 20,
  \end{align}
where $\circlearrowright$ represents another 24 terms to be included to ensure the cyclic symmetry of the Bell expression.

For $\IFull{LG_5}$, we have:
\begin{equation}
  \begin{split}
    \IFull{LG_5} := &E_5(1,3,\varnothing,\varnothing,\varnothing)+E_5(3,1,3,\varnothing,\varnothing)+E_5(\varnothing,3,1,3,\varnothing)\\
     + &E_5(2,2,3,\varnothing,\varnothing)+E_5(3,2,2,3,\varnothing) +E_5(1,\varnothing,1,3,\varnothing)\\
    +&E_5(1,3,3,1,3)+E_5(1,3,\varnothing,3,1)+E_5(3,2,1,1,2)\\
    +&E_5(3,1,\varnothing,1,3)-E_5(2,1,2,3,\varnothing)+E_5(2,2,\varnothing,1,3)\\
    +&E_5(2,2,3,3,1)+E_5(1,\varnothing,2,2,3)+E_5(1,\varnothing,1,\varnothing,1)\\
    -&E_5(3,2,1,2,3)-E_5(2,1,2,\varnothing,1)+E_5(2,2,\varnothing,2,2)\\
    -&E_5(2,1,1,1,2) +1+\leftrightarrow\,\,  \overset{\L}{\le} 20,
  \end{split}
\end{equation}
where $\leftrightarrow$ represents another 12 terms that have to be included to ensure that the Bell expression is invariant under the ``flip'' symmetry, i.e., the simultaneous permutation of party $j$ with party $n+1-j$ for all $j=1,2,\ldots, \lfloor\frac{n}{2}\rfloor$.

The $k$-producible bounds of these inequalities are summarized below in Table~\ref{tab:KProducibleBounds_IG_ThreeSettings}.

\begin{table}[h!]
\begin{center}
  \begin{tabular}{|c|c|c|c|c|c|} 
    \hline
  &1-prod.  & 2-prod. & 3-prod. & 4-prod. & 5-prod. \\ [0.5ex] 
    \hline\hline
    $\IFull{RG_3}$& 6 & $^\ddag2\sqrt{2}+4$ & *8 & - & - \\ 
    \hline
    $\IFull{{\rm RG_4}}$ & $12$ & $^\ddag12$ & $^\ddag12$ & *16 & - \\
    \hline
    $\IFull{{\rm RG_5}}$ &20 & $^\ddag20$ & $^\ddag20$ &  $^\ddag21.8564$ & *32  \\
    \hline \hline
    $\IFull{{\rm LG_5}}$& 20 & ? & ?  & ? & *32\\ 
    \hline
\end{tabular}
\end{center}
\caption{\label{tab:KProducibleBounds_IG_ThreeSettings} Summary of \emph{upper} bounds  on the quantum $k$-producible bounds of $\IFull{RG_n}$ and $\IFull{LG_n}$ for $n$ up to 5. The entries marked with ``?" are those where we have not succeeded (so far) in obtaining a reliable upper bound by solving the corresponding semidefinite programs.}
\end{table}

\bibliography{DIWED}
\end{document}